\documentclass[pre,twocolumns,amsmath,amssymb,showpacs,floatfix]{revtex4}

\usepackage{epsfig,amssymb,amsmath,bm,graphicx}

\makeatletter
\def\graphicscale{\twocolumn@sw{0.33}{0.4}}
\makeatother

\def\spose#1{\hbox to 0pt{#1\hss}}
\def\lesssim{\mathrel{\spose{\lower 3pt\hbox{$\mathchar"218$}}
 \raise 2.0pt\hbox{$\mathchar"13C$}}}
\def\gtrsim{\mathrel{\spose{\lower 3pt\hbox{$\mathchar"218$}}
 \raise 2.0pt\hbox{$\mathchar"13E$}}}

\def\<{\langle}
\def\>{\rangle}

\begin{document}

\title{Polymer models with optimal good-solvent behavior}

\author{G. D'Adamo$^1$, A. Pelissetto$^2$}

\affiliation{$^1$ SISSA, International School for Advanced Studies, via Bonomea 265, I-34136 Trieste, Italy \\
$^2$Dipartimento di Fisica, Sapienza Universit\`a di Roma and
INFN, Sezione di Roma I, P.le Aldo Moro 2, I-00185 Roma, Italy}

\begin{abstract}
We consider three different continuum polymer models, that all depend on 
a tunable parameter $r$ that determines the strength of the 
excluded-volume interactions.  In the first model chains 
are obtained by 
concatenating hard spherocylinders of height $b$ and diameter $rb$
(we call them thick self-avoiding chains).
The other two models are generalizations of the tangent hard-sphere and of 
the Kremer-Grest models. We show that, for a specific value $r^*$, 
all models show an optimal behavior: asymptotic long-chain behavior is 
observed for relatively short chains. For $r < r^*$, instead, 
the behavior can be parametrized
by using the two-parameter model that also describes the thermal crossover
close to the $\theta$ point. The bonds of thick self-avoiding chains
cannot cross each other and, therefore, the model is suited for 
the investigation of topological properties and for dynamical studies.
Such a model also provides a coarse-grained description of double-stranded DNA, 
so that we can use our results to discuss under which conditions DNA can be 
considered as a model good-solvent polymer.
\end{abstract}

\pacs{05.20.Jj, 05.20.Gg, 05.70.Ce, 65.20.De}
%% 65.20.De	General theory of thermodynamic properties of liquids,
%%                including computer simulation
%% 05.20.Gg	Classical ensemble theory
%% 05.20.Jj	Statistical mechanics of classical fluids
%% 05.70.Ce	Thermodynamic functions and equations of state

\maketitle

\section{Introduction}

In the last decades, a significant advancement in the theoretical 
understanding of polymeric systems has been possible thanks to the 
combination of simulative approaches \cite{Binder-96}, scaling arguments, 
and renormalization-group calculations
\cite{deGennes-79,Freed-87,DE-88,dCJ-book,Schaefer-99}.
Most of the large-scale dynamic and static properties 
of synthetic homopolymers over a wide concentration range have been 
shown to be universal: the predictions of \emph{coarse-grained} polymer models,
often with only little connection to realistic systems, describe
quite accurately the extensive experimental data  
collected from scattering, osmometric, and rheological experiments on 
chemically very different polymer solutions.
This very successful approach has been applied to homopolymers of different
topology and also extended to
biopolymers; see, e.g., \cite{VFDLRRD-05,NYSK-13,TMDD-13}.

The physical appeal of the universal scaling picture  relies on the 
possibility of formulating predictions invoking only a limited 
number of \emph{explanatory microscopic variables} often connected to, and 
hence inferable from, experimentally accessible properties. For instance, 
the thermal crossover observed in dilute polymer solutions is typically 
parametrized using only a suitable combination of a microscopic excluded-volume 
parameter, expressing the average strength of the solvent-mediated 
monomer-monomer interaction, and the degree of polymerization
\cite{deGennes-79,Freed-87,dCJ-book,Schaefer-99}.
This absence of tight  constraints in the physical schematization of 
polymers has determined over the years the use, both in theory and simulations,
of widely different  coarse-grained chain models. Popular instances  
are the lattice self-avoiding walk model, the tangent hard-sphere (sometimes
also named pearl-necklace) model
\cite{DH-86,DH-88} and the Kremer-Grest \cite{GK-86} model. 
It is important to stress that only the large-scale behavior is 
universal, i.e., the behavior on length scales of the order of the radius
of gyration $R_g$ or of de Gennes correlation length $\xi$
\cite{deGennes-79}, when they are significantly larger than any 
microscopic scale. Therefore, universality holds for long polymers in the 
dilute and semidilute regimes, but not for melts, in which 
the characteristic polymer length scales are in the microscopic domain
(although in this case some scaling laws still hold). 
The universal behavior observed in 
 the dilute and semidilute regimes can be rationalized using 
renormalization-group arguments, thanks to the de Gennes mapping
\cite{deGennes-72,deGennes-79,dCJ-book} 
of a polymer system onto a spin model, whose critical behavior is by now 
very well understood \cite{Wegner-76,PHA-91,PV-02}.

While universality is well established for the thermodynamic behavior of 
polymer solutions, the universality of the polymer dynamics is less clear. 
On general principles, one would expect a universal behavior on large
time scales, i.e., for $t \gg \tau_0$, where $\tau_0$ is the Kuhn monomer 
relaxation time, which can be understood using the standard
renormalization-group tools \cite{Oono-85,WDF-86,WF-86,SFO-85,SH-89}.
However, any physical dynamics should include the condition of 
bond-noncrossability, which represents a nonlocal dynamical constraint. 
In the dilute regime, in which different polymers do not overlap, this
constraint should not be crucial and thus universality is expected to hold. 
In the semidilute regime, the role of the constraint is less clear 
as polymers strongly overlap, although there is no entanglement as the 
monomer density is vanishingly small. 

It must be noted that,
although all models exhibit the same asymptotic
behavior for a large number of monomeric units,  when
relatively short chains are used, they deviate systematically from each other, 
leading to a serious issue of interpretability of the results. 
A brute-force numerical solution of the problem 
consists in simulating long chains, which 
can be done quite efficiently for single isolated chains using clever 
algorithms developed in the years, see, e.g., 
\cite{Grassberger-97,Clisby-10-JSP}. This is not feasible 
in finite-density simulations, in which typically
the chain length never exceeds a few hundred 
monomeric units, making the quantitative comparison of different models 
rather difficult. For this reason, in the last years models rapidly 
approaching the universal large degree-of-polymerization limit have been 
proposed, see, for instance,
\cite{BN-97,LK-99,CMP-06,DP-16,Clisby-17}.
Alternatively, a number of first-principle coarse-grained approaches 
have been devised
\cite{BLHM-01,MullerPlathe-02,PCH-07,PK-09,DPP-12-Soft,DPP-12-JCP,%%
KVMP-12,Noid-13},  which reduce the complexity of the system by integrating 
out the short-scale degrees of freedom. Unfortunately, all these 
proposals, because of the lack of bond non-crossability, cannot be easily 
employed in  dynamical studies (see \cite{PB-01} for an 
algorithmic way out), 
in which the topology or the concatenation properties 
of a chain must not change under any physical local dynamics. 
Beside the determination of polymer dynamical properties, this also represents 
a serious issue  when looking at the knotting properties 
of macromolecules, which have  gained a considerable attention in the last 
years \cite{OW-07,FSS-11,VRSWW-92,TRFM-13,EB-14,WC-86,VR-09}.

In this paper we wish to develop continuum polymer models that 
rapidly approach the universal scaling limit and that can, therefore,
be used to investigate the thermodynamic properties of 
polymer solutions with a limited computational effort.
It is important to note that the optimal interaction is obtained by studying 
the length dependence of specific observables for {\em linear} chains. However, 
the universality of subleading-amplitude ratios, which has been extensively
verified in the context of spin systems \cite{AA-80,PV-02,PHA-91}, 
indicates that the cancellation of 
the leading finite-length corrections only depends
on the specific nature of the 
interactions, while it is independent of the specific observable one is 
considering. Moreover, extensive numerical studies show that these
finite-length corrections are not 
related to the polymer topology, and are always 
controlled by the same renormalization-group operator; for instance, the length 
dependence is specified by the same exponent
$\Delta$ \cite{JRWM-90,FZ-11,HNG-04,RP-13}, 
which takes the value \cite{Clisby-10} $\Delta \approx 0.53$. 
Therefore, an optimal interaction for linear chains is also optimal when
considering other types of polymer conformations. For star polymers 
this was explicitly verified in \cite{HNG-04,RP-13}.
It should be noted that, 
for cyclic chains, renormalization-group theory guarantees the optimality of
the models only if averages are taken over all chains, independently of 
the knot type.  If one considers instead
polymers of fixed knot type, the behavior is less clear, as new
corrections might appear. This issue deserves additional investigations.
The models we consider are also relevant for studies of mixtures of 
polymers and other nanoparticles, although in this case a fast convergence 
is only obtained by additionaly tuning the polymer-nanoparticle 
interactions \cite{DP-16}.

We will discuss three different models. 
First, we analyze the thick self-avoiding chain
\cite{VLKFKC-92,RCV-93,VC-95,GM-99,SM-00,MPR-08,UD-15-17,PZC-16},
which has been studied at length in the past as a 
minimalistic model for double-stranded DNA filaments
\cite{VLKFKC-92,RCV-93,VC-95}.
Specifically, we determine the optimal thickness for which, under 
good-solvent (GS) conditions, the universal large degree-of-polymerization 
limit can be obtained for relatively short chains. As a byproduct of the 
calculation, we 
also determine the thickness crossover of the model, characterizing 
quantitatively the   
region  between the ideal-chain limit and the GS behavior. 
The resulting behavior can be considered as representative of the one 
observed for DNA chains under various electrostatic screening conditions
\cite{RCV-93},
at least in those cases in which the size of the chain is somewhat 
larger than the persistence length. 

We also consider two popular models, often used to study the behavior of 
polymer solutions, the tangent hard-sphere model
\cite{DH-86,DH-88} and the Kremer-Grest \cite{GK-86} model. We show that 
both of them exhibit very strong scaling corrections, so that results depend
significantly on the number of monomers used. Both models can be easily
generalized so that, by tuning a single parameter, one can obtain an optimal
model, which approaches the asymptotic limit for small contour lengths.
However, in the resulting optimal models bonds can cross each other.
Therefore, they cannot be straightforwardly 
used in dynamical studies and whenever topology and 
concatenation  are important. 

The paper is organized as follows. In Sec.~\ref{sec2} we define the 
thick self-avoiding chains and discuss their relation with 
the models used in the description of double-stranded DNA and in 
the mathematical knot literature
\cite{VLKFKC-92,RCV-93,VC-95,GM-99,MPR-08}.
The thickness crossover and the behavior close to the GS regime
are analyzed in Secs.~\ref{sec3} and \ref{sec4}, respectively.
In Ref.~\ref{sec5} we extend the previous results to the tangent hard-sphere
\cite{DH-86,DH-88} and to the Kremer-Grest \cite{GK-86} models. Finally,
in Sec.~\ref{sec6} we draw our conclusions. In the Appendix
we compute the asymptotic expansion of an integral useful for the 
discussion of the Kremer-Grest model.

\section{Thick self-avoiding chains} \label{sec2}

\begin{figure*}[tb!]
\begin{center}
\begin{tabular}{c}
\includegraphics[width=0.5\textwidth,angle=0]{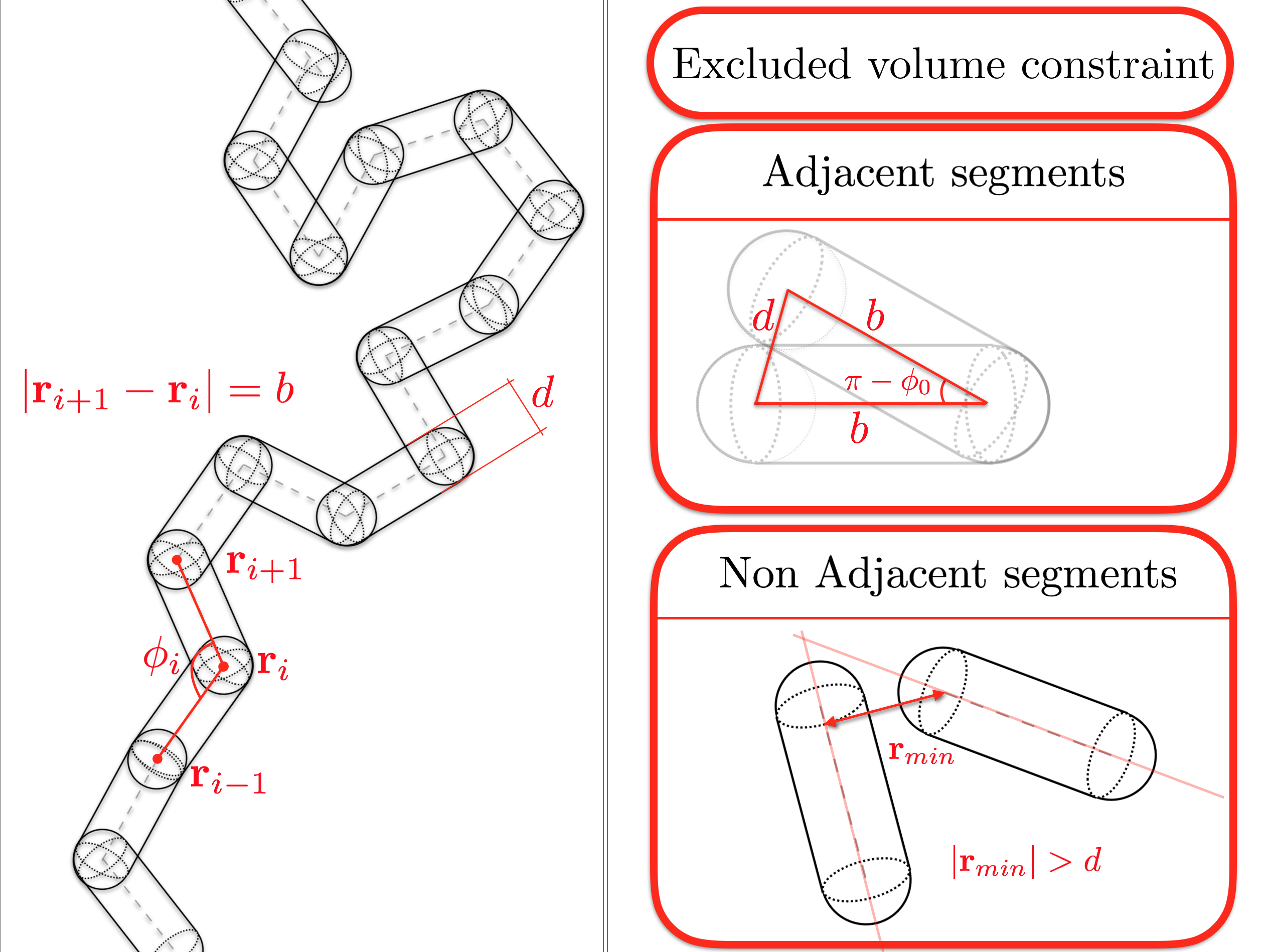} \\
\end{tabular}
\end{center}
\caption{The model. On the left, we show a typical chain of spherocylinders.
We require that nonadjacent segments do not overlap, so $r_{\rm min} > d$,
where $r_{\rm min}$ is the minimum distance between the axes of the two
cylinders. Adjacent spherocylinders overlap, but the excluded-volume 
condition implies $\phi_i < \phi_0 = \pi - 2 \arccos (d/2b)$.}
\label{cartoon}
\end{figure*}

We consider here a polymer model that has often been used to describe 
double-stranded DNA at a coarse-grained level \cite{VLKFKC-92,RCV-93,VC-95}.
A polymer with $L$ monomers is modelled as 
a chain made by $L-1$ spherocylinders, see Fig.~\ref{cartoon}. 
Specifically, a chain is defined by
$L$ points $\{{\bm r}_1,\ldots,{\bm r}_L\}$ such that 
the distance between subsequent units is fixed, i.e.,
$|{\bm r}_i - {\bm r}_{i-1}| = b$. 
Excluded volume effects are accounted for  by treating each segment
connecting 
two successive points as the axis of a spherocylinder, i.e., of 
a hard cylinder of 
height $b$ and diameter (thickness) $d$, capped by two half spheres of
radius $d/2$. Because of their steric encumbrance, 
nonadjacent spherocylinders cannot overlap.
This condition is easily verified numerically, checking whether 
the minimum distance 
$r_{\rm min}$ between the axes of the two cylinders satisfies 
$r_{\rm min} > d$. Such distance can be computed using the fast algorithm 
of \cite{VL-94}.
Successive spherocylinders can instead partly overlap. 
In the limit $d\to 0$ the excluded-volume constraint disappears and we obtain 
a freely-jointed chain. 

The model we consider is very similar to that considered in the mathematical
knot literature, see, e.g., \cite{GM-99,MPR-08,PZC-16}. 
The main difference
concerns the local constraint between two successive units. 
A model that generalizes ours and that of 
\cite{GM-99,MPR-08,PZC-16} is 
obtained as follows. We define 
${\bm e}_i = {\bm r}_{i+1} - {\bm r}_i$ and the angle $\phi_i$
($0\le \phi_i \le \pi$) as 
\begin{equation}
\cos \phi_i = {{\bm e}_i \cdot {\bm e}_{i+1} \over 
               |{\bm e}_i| |{\bm e}_{i+1}| }.
\label{def-phi}
\end{equation}
A more general model is obtained by
requiring $\phi_i < \phi_0$, for some fixed $\phi_0$.
In our model, the excluded volume requirement for nonadjacent spherocylinders 
implies $|{\bm r}_i - {\bm r}_{i-2}| > d$, so we 
have $\phi_0 = \pi - 2 \arccos (d/2b)$. 
The model of \cite{GM-99,MPR-08,PZC-16} differs only in the 
value of $\phi_0$: they take $\phi_0 = \pi - 2 \arccos (d/b)$. 

Thick self-avoiding chains provide a coarse-grained model for 
double-stranded DNA, 
where $b$ may be identified with the Kuhn length \cite{RC-libro}
$l_K$ ($l_K$ is twice the 
persistence length), while 
$d$ is an effective thickness, that depends on the amount of salt (pH) 
present in the solution \cite{RCV-93,VC-95}.
In principle, 
by tuning the angle $\phi_0$ defined above, we can also obtain a more 
accurate description in which $b < l_K$ and each spherocylinder represents
a shorter DNA segment. Alternatively, one can introduce a bending energy
$E_b = \alpha \sum_i (1 - \cos \phi_i)$, as 
in \cite{VLKFKC-92,RCV-93,VC-95}, depending on a bending 
parameter $\alpha$.

Depending on the value of $d$, finite-length polymers can exhibit a different
behavior. For small values of the thickness $d$, i.e., for very small excluded-volume interactions,
the behavior is very similar to that of $\theta$-point chains. As $d$
increases,
excluded-volume interactions become more effective, and therefore the behavior
becomes gradually closer to that expected for a GS polymer.
Therefore, the thickness $d$
plays the same role as that of the temperature $T$ in the 
analysis of the thermal crossover between the $\theta$ and the GS
regime of homopolymer
solutions. In that context, solvent quality is usually parameterized by using 
the Zimm-Stockmayer-Fixmann variable \cite{ZSF-53} 
$z \sim (T- T_\theta) L^{1/2}$, where $T_\theta$ is the $\theta$ temperature. 
The $\theta$ behavior is observed for $z\approx 0$, while the GS regime is 
reached in the limit $z\to \infty$. 
We will show here that an analogous 
quantity can be defined for thick chains, replacing the temperature difference 
with an appropriate function of $d$. This allows us to map the crossover 
due to the change of the steric encumbrance on 
the standard thermal crossover, for which a wealth of theoretical and 
numerical results are 
available \cite{Yamakawa-71,Schaefer-99,dCJ-book,CMP-08,DPP-13-Thermal}.

To study the crossover, we perform Monte Carlo simulations 
of isolated chains for values of $d/b$ ranging from 0.0125 to 0.7
and polymer lengths $L$ from $250$ to $8000$. 
We use a continuum generalization of the pivot algorithm \cite{MS-88},
implementing the Kennedy algorithm \cite{Kennedy-02}
to speed up the self-intersection check.
We compute the second virial coefficient\footnote{
The coefficient $B_2$ is defined from the expansion of the
polymer (osmotic) pressure $P$ is powers of the number density
$\rho = N/V$, where $N$ is the number of polymers in a volume $V$.
For $\rho\to 0$, we have
$\beta P/\rho = 1 + B_2 \rho + O(\rho^2)$.}
$B_2$ and the radius of gyration $R_g$.
Then, we consider the  
universal combination $A_2 = B_2 R_g^{-3}$, which vanishes in the  
noninteracting case ($d=0$) and takes the value \cite{CMP-06,DP-16}
$A_{2,GS} \approx 5.501$ for polymers under GS conditions. 
The interpenetration ratio $\Psi$ often used in the experimental literature 
is defined \cite{Yamakawa-71,Schaefer-99,dCJ-book}
as $\Psi = {1\over4} \pi^{-3/2} A_2$. We also consider the 
expansion (swelling) ratio $\alpha^2_g$ defined as the ratio between $R_g^2$
and the corresponding value for a freely jointed (ideal) chain of $L$ sites:
\begin{equation}
R^2_{g,0}=\frac{Lb^2}{6}\left(1-\frac{1}{L^2}\right).
\end{equation}

\section{Thickness crossover} \label{sec3}

\begin{figure}[tb!]
\begin{center}
\begin{tabular}{c}
\includegraphics[width=0.45\textwidth]{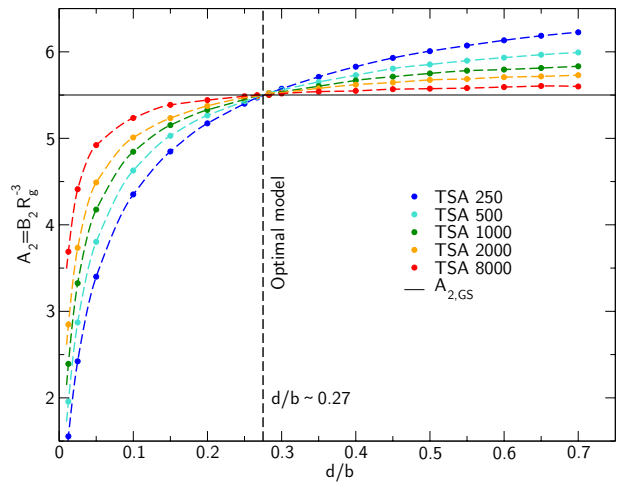} \\
\end{tabular}
\end{center}
\caption{Second-virial combination $A_2$ vs $d/b$ 
for several values of the chain length $L$. The horizontal line corresponds 
to the good-solvent value $A_{2,GS} = 5.501$. }
\label{fig:A2vsd}
\end{figure}

\begin{figure}[tb!]
\begin{center}
\begin{tabular}{c}
\includegraphics[width=0.45\textwidth]{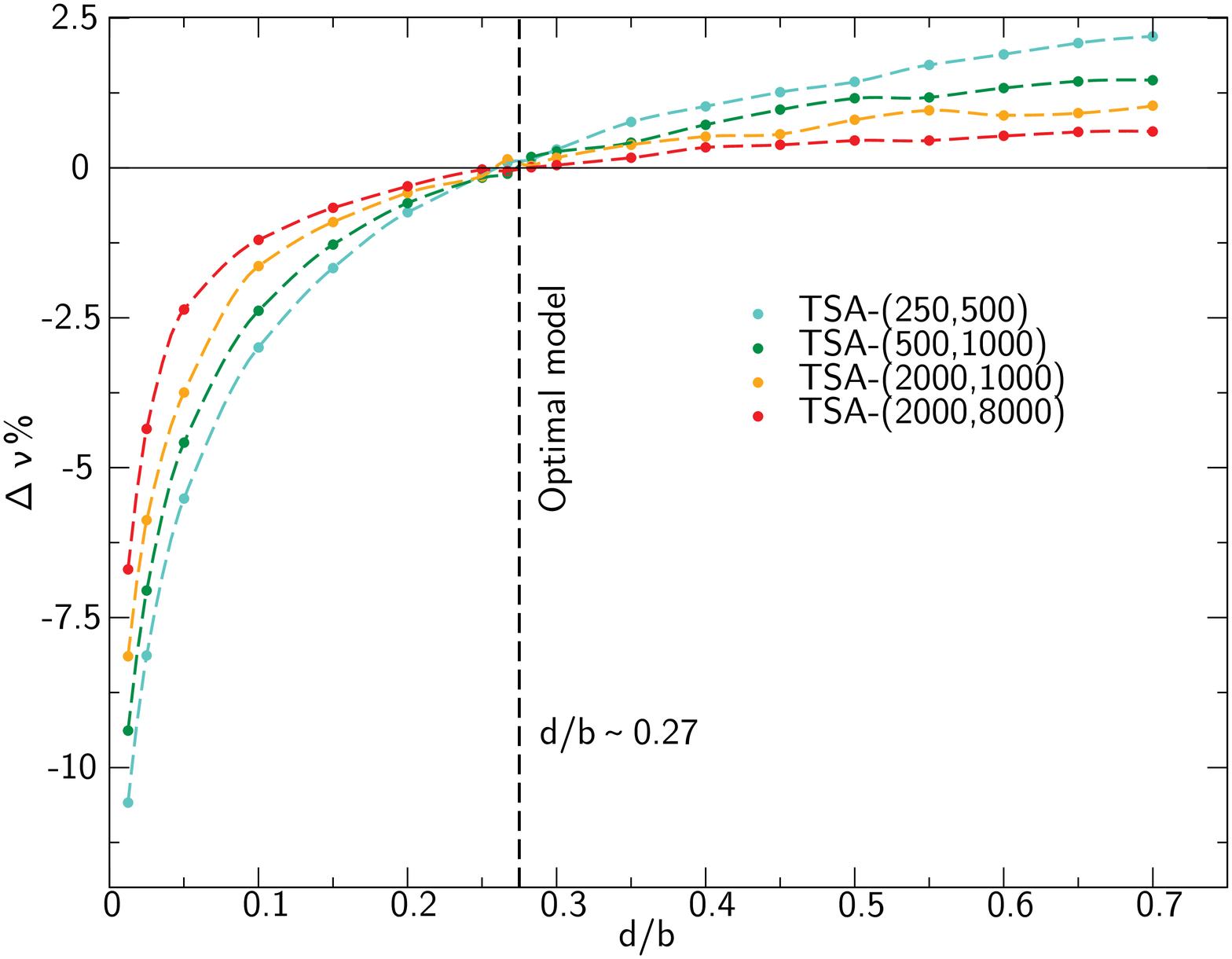} \\
\end{tabular}
\end{center}
\caption{Plot of $\Delta \nu = \nu_{\rm eff} - \nu_{GS}$, 
for several values of the chain lengths ($L_1$, $L_2$). 
}
\label{fig:nueff-TSA}
\end{figure}

In Fig.~\ref{fig:A2vsd} we report the second-virial combination $A_2$ 
versus the 
thickness $d$. The observed behavior is completely analogous to that observed 
in experimental systems (see, e.g., \cite{Berry-66} and 
the recent reanalysis
presented in the supplementary material of 
\cite{DPP-14-GFVT}) or in Monte Carlo simulations of 
polymer models \cite{PH-05}, once we replace $d/b$ with $T-T_\theta$. 
At fixed $d\not=0$ the estimates of $A_2$ converge to the 
GS value as $L\to \infty$, with discrepancies that increase 
as $d$ decreases towards zero. For $d/b \approx 0.27$ 
we obtain GS behavior even for small values of $L$. 
For $d/b\lesssim 0.27$ the approach to the GS value 
is from below, while for larger values 
of $d/b$ the finite-$L$ estimates are larger than the asymptotic value. 
Finite-length corrections to the GS behavior are therefore 
both positive and negative,
a phenomenon well documented in the thermal case \cite{Schaefer-99,KS-94}.

It is also interesting to discuss 
how the size of the chain changes with $L$ for different
values of $d/b$. For this purpose, we define an effective Flory exponent 
$\nu_{\rm eff}$,
\begin{equation}
\nu_{\rm eff}(d/b;L_1,L_2) = {\ln [R_g(d/b;L_1)/R_g(d/b;L_2)] \over 
                              \ln (L_1/L_2)}.
\end{equation}
In the good-solvent limit we should find $\nu_{\rm eff} = \nu_{GS}$,
where $\nu_{GS}$ is the universal good-solvent exponent, which is 
known with high accuracy
\cite{Clisby-10}: $\nu_{GS} = 0.587597(7)$.
In Fig.~\ref{fig:nueff-TSA} we report $\Delta \nu = \nu_{\rm eff} - \nu_{GS}$ 
for several pairs $(L_1,L_2)$. The results are analogous to those 
shown in Fig.~\ref{fig:A2vsd}. For $d/b < 0.27$, the effective exponent is 
smaller than $\nu_{GS}$ and increases toward the good-solvent value as 
$L_1$ and $L_2$ increase. For $d/b > 0.27$ the opposite occurs, while 
for $d/b\approx 0.27$, the radius of gyration scales quite 
precisely as $L^{\nu_{GS}}$.

We will now show that the crossover for $d/b\lesssim 0.27$, 
interpolating between
ideal and GS behavior, can be parametrized by using the standard 
two-parameter model (TPM) scaling functions
\cite{Yamakawa-71,dCJ-book,Schaefer-99}. 
For this purpose we define the scaling variable $z$ as 
$z = a v(d/b) L^{1/2}$, 
where the excluded-volume parameter $v(d/b)$ is identified with \cite{Hall-80} 
the adimensional microscopic virial coefficient---the one 
associated with the interaction of two 
isolated monomers. For pairs of identical spherocylinders we have 
\cite{Isihara-50,IH-51,Kihara-53} 
\begin{equation}
v(d/b) = 
\frac{2\pi}{3} {d^3\over b^3} \left(1+\frac{3b}{2d}+ \frac{3b^2}{8d^2}\right).
\end{equation}
The constant $a$ is fixed by the convention that 
$\Psi \approx z$ for $z \to 0$.

In order to determine $a$, we performed simulations for small values of $z$ 
for $L=1000$, corresponding to values of $A_2$ in the interval 0.5-1.5. 
We finally fitted the results (they are reported in Table~\ref{Table-smallz})
to the accurate TPM expression of \cite{CMP-08}:
\begin{eqnarray}
A_2(z)&=&4 \pi ^{3/2} z f_2(z)^{-1/4} 
\label{A2TPM}\\
f_2(z) &=&   1 + \nonumber \\
&& 268.96 z^4+331.99 z^3+126.783 z^2+19.1187 z.
\nonumber 
\end{eqnarray}
We find $a\approx 0.647(2)$. As a consistency check we can compare 
the estimated $\alpha_g$ with the TPM prediction \cite{CMP-08},
\begin{equation}
\alpha_g(z) = (1 + 10.9288z + 35.1869 z^2 + 30.4463 z^3)^{0.0583867}. 
\label{alphagTPM}
\end{equation}
The results are reported in Table~\ref{Table-smallz}: deviations are small,
of less than approximately 1\%.

\begin{table}[t!]
\caption{Monte Carlo estimates of $A_2$ and $\alpha^2_{g}$ for several values
of $d$ and $L=1000$. For each $d/b$ and $L$ we compute $z$ using $z = 0.647
v(d/b) L^{1/2}$ and report the corresponding TPM predictions [we use 
Eqs.~(\ref{A2TPM}) and (\ref{alphagTPM})]. 
}
\label{Table-smallz}
\begin{center}
\begin{tabular}{cccccc}
\hline
\hline
\footnotesize
$d/b$  & $z$ & $A_2$ & $A_{2,TPM}$ & $\alpha^2_g$ & $\alpha^2_{g,TPM}$ \\
\hline
0.00160& 0.02582& 0.516(1)& 0.513&  1.0302(3)  &  1.032   \\
0.00253& 0.04105& 0.772(1)& 0.767 & 1.0463(3)  &  1.049\\
0.00358& 0.05838 &1.024(1)& 1.022 & 1.0640(3)  &  1.069\\
0.00476& 0.07796& 1.275(2)& 1.276 & 1.0825(3)  &  1.089\\
0.00609& 0.10031& 1.524(2)& 1.529 & 1.1036(3)  &  1.112\\
\hline
\hline
\end{tabular}
\end{center}
\end{table}

\begin{figure}[tb!]
\begin{center}
\begin{tabular}{c}
\includegraphics[width=0.45\textwidth]{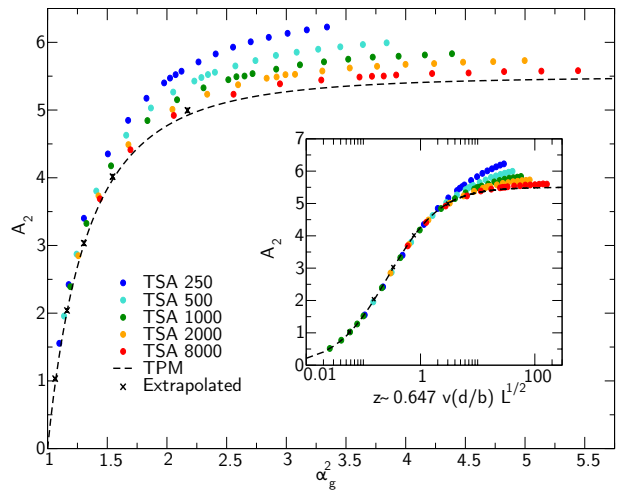} \\
\end{tabular}
\end{center}
\caption{Second virial combination $A_2$  versus
$\alpha^2_g$. We report the numerical results for several values
of $L$, the extrapolations (see Table~\ref{tab:suppl-SP-cross})
of data at fixed $z$ and the TPM predictions from \cite{CMP-08}. 
In the inset we report $A_2$ versus $z = 0.647 v(d/b) L^{1/2}$.
}
\label{fig:A2-cross}
\end{figure}

In Fig.~\ref{fig:A2-cross} we compare the TPM crossover curves with the whole
set of Monte Carlo data. The results for $A_2$ are well reproduced up to 
$A_2 \approx 5$, while larger discrepancies are observed for the swelling
ratio, a phenomenon already observed in the analysis of the experimental
data \cite{Berry-66,DPP-14-GFVT}, which is due to the fact that the 
corrections to the GS behavior increase as $z\to \infty$.

\begin{center}
\begin{table*}[tb!]
\caption{Crossover behavior in the spherocylinder model. We perform runs at
values of $d/b$ such that $A_{2,TPM}(z) = 1,2,3,4,5$, where
$A_{2,TPM}(z)$ is given in Eq.~(\ref{A2TPM}) and
$z = 0.647 v(d) L^{1/2}$. We report the estimates of  $A_2$ and $\alpha_g^2$
for several values of $L$ and the corresponding large-$L$
extrapolation (we assume \cite{DJ-72,CMP-08,BN-97}
$A_{2,L} = A_{2,\rm extr} + b_1 L^{-1/2} + b_2/L$).
Finally $\Delta = 100 (1 - B_{\rm Extr}/B_{TPM})$, $B = A_2$ or $\alpha_g^2$.
}
\label{tab:suppl-SP-cross}
\begin{tabular}{c|cccccccc}
\hline
\hline
\footnotesize
$z$ & $A_{2,TPM}(z)$ & 250 & 500 & 1000 & 2000 & 4000 & Extr. & $\Delta\%$ \\
\hline
0.056778 & 1 &0.990(1) &0.992(2)& 1.005(4)& 1.002(8)& 1.010(9)&1.03(2)& 3\\
0.151493 & 2 &1.958(2) &1.972(3)& 1.994(6)& 2.00(1) & 2.00(1) &2.04(2)& 2\\
0.331075 & 3 &2.929(2) &2.936(2)& 2.962(6)& 2.97(1) & 2.97(2) &3.03(3)& 1.1\\
0.767908 & 4 &3.934(3) &3.931(5)& 3.944(8)& 3.94(1) & 3.96(2) &4.02(3)& 0.4\\
3.01381  & 5 &5.154(3) &5.045(5)& 5.009(9)& 4.97(2) & 4.98(2) &4.99(3)& 0.1\\
\hline
$z$ & $\alpha^2_{g,TPM}(z)$ & 250 & 500 & 1000 & 2000 & 4000 & Extr. \\
\hline
0.056778&1.0668&1.0572(2)&1.0604(4)&1.0616(7)&1.065(1) & 1.061(2) & 1.063(3) &
0.35\\
0.151493&1.1602&1.1314(2)&1.1395(4)&1.1461(7)&1.149(1) & 1.155(2) & 1.1625(25)&
  0.2 \\
0.331075&1.3020&1.2356(2)&1.2522(4)&1.2663(7)&1.276(1) & 1.283(2) &1.304(3) &
0.15 \\
0.767908&1.5553&1.4028(2)&1.4364(4)&1.4627(7)&1.4859(1)& 1.503(2) &1.544(3) &
0.7\\
3.01381&2.2857&1.8237(3)&1.8913(5)&1.9538(9)&2.010(2)  & 2.057(2) & 2.172(5)
& 5\\
\hline
\hline
\end{tabular}
\end{table*}
\end{center}

A more detailed check is presented in Table~\ref{tab:suppl-SP-cross}. Here 
we consider five different values of $z$ such that $A_{2,TPM}(z)$ is 
1,2,3,4,5, respectively, and several values of $L$. For each $z$ and $L$, 
we compute the corresponding value of $d/b$. Then, we perform simulations
of chains of length $L$ with the computed $d/b$. As it can be seen from 
Table~\ref{tab:suppl-SP-cross}, results show only a tiny dependence on $L$. 
Moreover, the extrapolated 
large-$L$ values are consistent with the TPM predictions, both for $A_2$ and for
the swelling ratio. Only in one case, the result for $\alpha^2_g$ at 
$z = 3.01381$, do we observe a significant difference. This may be due to
additional corrections to scaling that are not taken into account in the 
extrapolation (if we do not use the result for $L=250$, the extrapolated
value becomes 2.190(7), which is closer to the TPM estimate) and to 
the inaccuracy of Eq.~(\ref{alphagTPM})---the interpolation should be 
accurate at the level of a few percent.

It is important to stress that, if we change the model by considering 
a different value of $\phi_0$ (the parameter that controls the interaction of
two successive spherocylinders) or by adding a bending energy 
as in DNA models, only the constant $a$ 
entering the definition of $z$ changes. 
The TPM describes the crossover behavior of any large-scale quantity
for any model parameter, 
provided the constant $a$ is appropriately chosen.

\begin{figure}[tb!]
\begin{center}
\begin{tabular}{c}
\includegraphics[width=0.45\textwidth]{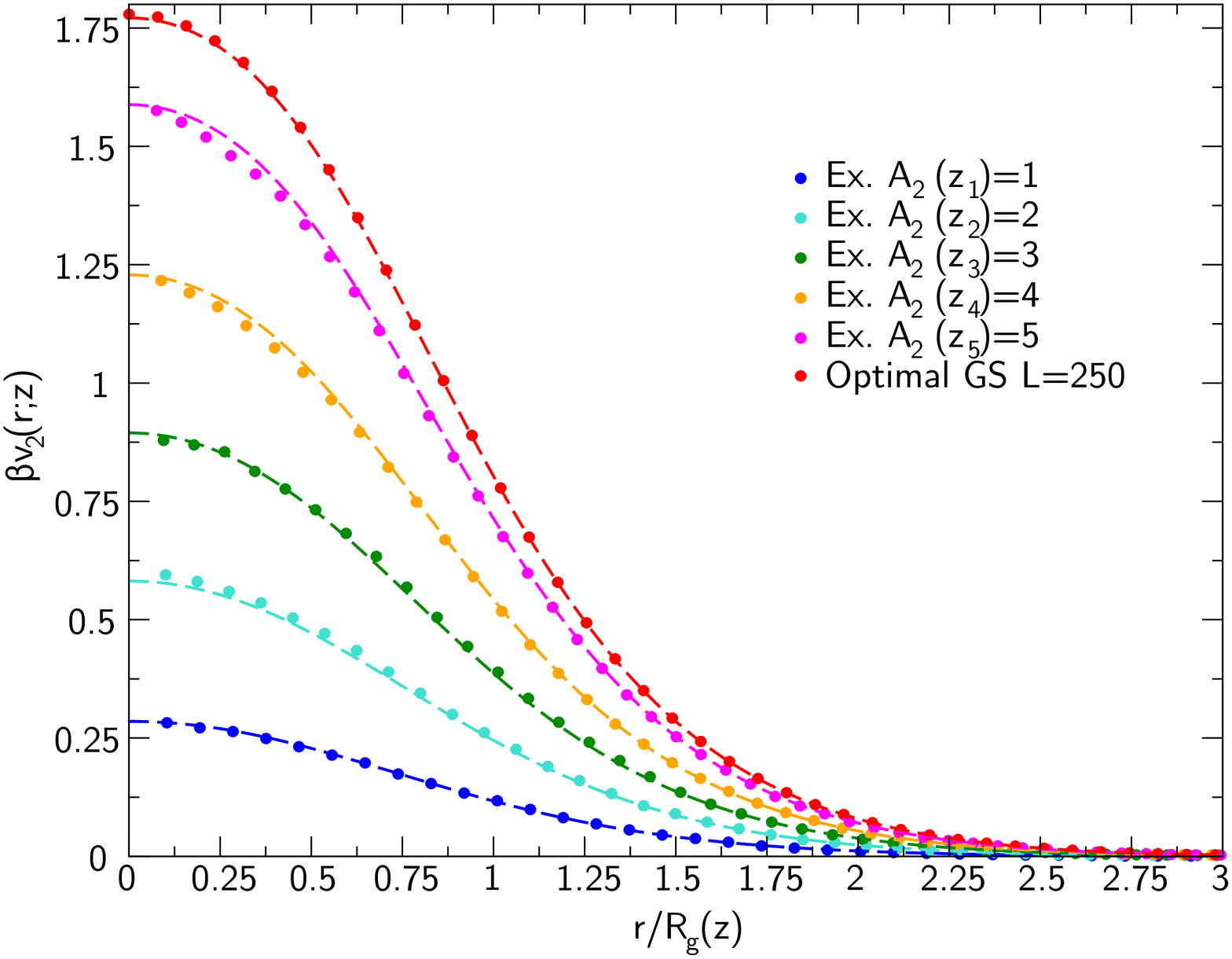} \\
\end{tabular}
\end{center}
\caption{Estimates of the two-body potential as a function of $r/R_g$. 
We report Monte Carlo results (points) for spherocylinder chains with 
$L=4000$ monomers at values of $d/b$ such that $A_2(z_n) = n$ and 
the results (lines) of \cite{DPP-13-Thermal} obtained by
using the lattice Domb-Joyce model. Data labelled ``Optimal GS L=250'' (points)
correspond to simulations of $L=250$ chains for 
$d/b = 0.275$ (the optimal model discussed in Sec.~\ref{sec4}). 
The line that goes through these data was obtained in \cite{PH-05},
by extrapolating self-avoiding walk results.
}
\label{fig:Vstar}
\end{figure}

\begin{figure}[tb!]
\begin{center}
\begin{tabular}{c}
\includegraphics[width=0.45\textwidth]{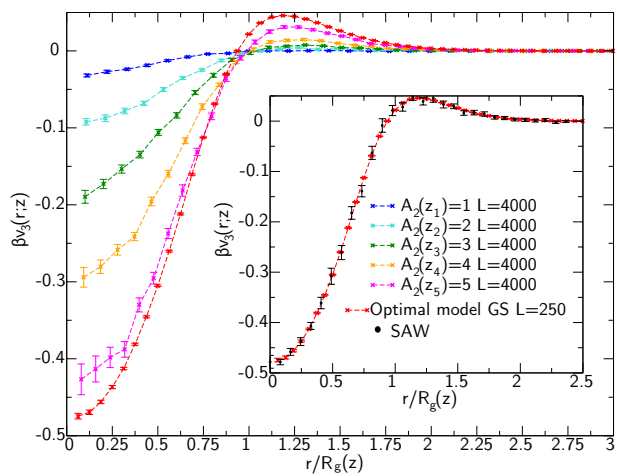} \\
\end{tabular}
\end{center}
\caption{Estimates 
of the three-body potential for equilateral configurations for the 
same values of $d/b$ and $L$ as in Fig.~\ref{fig:Vstar}. 
In the main frame, the lines going through the points are only meant to
guide the eye.
The data labelled ``SAW" in the inset were obtained in 
\cite{DPP-compressible} by using the lattice Domb-Joyce model. 
}
\label{fig:V3star}
\end{figure}

To verify the correctness of our parametrization of the steric crossover,
we consider two other quantities, which characterize the intermolecular
structure of a polymer solution in the dilute regime. 
We compute the effective 
center-of-mass two-body and three-body potential \cite{BLHM-01,Louis}.
In Fig.~\ref{fig:Vstar} we report our results for five different 
values of $d/b$ chosen so that $A_{2,TPM}(z_n) = n$ (the same 
values of $z$ considered in Table~\ref{tab:suppl-SP-cross}) and compare them
with those obtained in \cite{DPP-13-Thermal}, using the 
lattice Domb-Joyce model \cite{DJ-72}. We observe a very
good agreement, confirming that we have correctly identified the 
scaling variable $z$. In Fig.~\ref{fig:V3star} we show the corresponding 
three-body center-of-mass potentials, which, as expected, converge to 
zero as $z$ goes to zero. Again, they agree with those obtained by using the 
lattice Domb-Joyce model \cite{DPP-compressible}.

\section{Good-solvent behavior and optimal model} \label{sec4}

In the previous section we have discussed the crossover behavior for small 
values of $d$, up to $d/b\approx 0.27$, where GS behavior is observed 
even for small values of $L$. We wish now to focus on the behavior close to
this value of $d$. In general, for large values of $L$, any generic 
large-scale adimensional quantity $A$, which depends on $L$ and $d$, 
behaves as 
\begin{equation}
    A(L,d) = A_{GS} + a_{1,A}(d)/L^\Delta + a_{2,A}(d)/L^{\Delta_2} + \ldots
\end{equation}
where $A_{GS}$ is universal, i.e., independent of $d$.
The exponent
$\Delta$ is also universal [simulations of self-avoiding walks give
\cite{Clisby-10} $\Delta = 0.528(12)$], and so is\footnote{
To be precise there are three possible correction terms that have
exponent close to 1, see \cite{PH-05} for a discussion.
Here, they are lumped together in a single effective correction term.}
$\Delta_2 \approx 1$. The amplitudes $a_{1,A}(d)$ and $a_{2,A}(d)$ depend
instead on the model. However, given two different observables
$A$ and $B$, the amplitude ratios $a_{1,A}(d)/a_{1,B}(d)$ are
model independent \cite{AA-80}.

Finite-length corrections represent 
the main obstacle for a precise determination of
the leading, universal behavior under GS conditions. However, 
one can exploit the model dependence
of the corrections to identify optimal models for which the
leading scaling corrections vanish. In our case, we wish to determine
the value $d^*$ of the thickness parameter, for which the leading scaling
corrections vanish.
Note that, if $a_{1,A}(d^*)=0$ for a given observable $A$, the universality
of the amplitude ratios implies that $a_{1,B}(d^*) = 0$ for any
other observable $B$.

\begin{figure}[tb!]
\begin{center}
\begin{tabular}{c}
\includegraphics[width=0.45\textwidth]{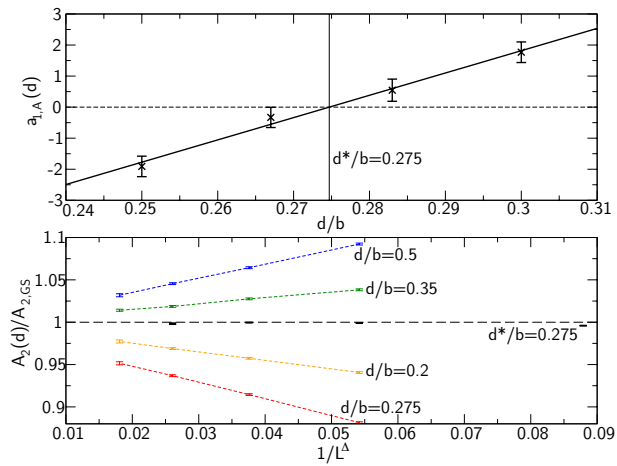} \\
\end{tabular}
\end{center}
\caption{Top: Scaling correction coefficient $a_1(d)$ versus $d/b$; the line 
is the linear interpolation. Bottom: $A_2(d,L)/A_{2,GS}$ versus $L^{-\Delta}$,
$\Delta = 0.528$, for several values of $d/b$. 
}
\label{fig:A2dstar}
\end{figure}

To determine $d^*$, we consider $A_2(d,L)$. We fit $A_{2}(d,L) - 
A_{2,GS}$ to 
\begin{equation}
   {a_1(d)\over L^\Delta} + {a_2(d)\over L^{\Delta_2}},
\label{fitA2}
\end{equation}
fixing $A_{2,GS} = 5.500$, $\Delta = 0.528$ and $\Delta_2 = 1$. 
We repeat the procedure for several values of $b$ in the interval 
$0.25 \le d/b \le 0.30$, obtaining estimates of  
$a_{1}(d)$ that are fitted to $a_1(d) = c (d - d^*)$. 
We obtain 
\begin{equation}
d^*/b = 0.275(2),
\end{equation}
where the error takes into account the variation of the estimates if 
$\Delta$ varies by \cite{Clisby-10} 0.012 and $\Delta_2$ by 0.1. 
We have then performed 
simulations for such a value of $d/b$. The results reported in 
Table~\ref{tab:dstar} completely confirm the analysis. 
It is important to note that the estimate of $d^*/b$ is strictly model 
dependent, and therefore a different result would be obtained if 
the model is changed by adding, for instance, a bending interaction term
or by considering a different value for the parameter $\phi_0$ that controls
the interaction between two adjacent spherocylinders.

\begin{table}[t!]
\caption{Second-virial combination $A_2$, asphericity $A_{\rm sph}$, 
ratios $L_{3,1}$ and $L_{2,1}$ 
between the eigenvalues of the gyration tensor, and ratio $A_{ge}$, 
for $d/b = 0.275$ and several values of $L$. 
The extrapolated ($L=\infty$) value of $A_2$ is taken from \cite{DP-16}, 
those for the shape factors are extrapolations of the 
results of \cite{Sciutto-96}, while the 
estimate of $A_{ge}$ is taken from \cite{Clisby-10}.
}
\label{tab:dstar}
\begin{center}
\begin{tabular}{c|ccccc}
\hline
\hline
\footnotesize
$L$       & $A_{2}$  & $A_{\rm sph}$   &  $L_{3,1}$   & $L_{2,1}$  
    & $A_{ge}$ \\
\hline
100   & 5.478(1)  & 0.42962(3) & 13.821(2) & 2.9222(4)  & 0.16093(6)\\
250   & 5.494(1)  & 0.43054(3) & 13.908(2) & 2.9351(5)  & 0.16020(6)\\
500   & 5.497(2)  & 0.43062(5) & 13.925(3) & 2.9393(6)  & 0.16006(6)\\
1000  & 5.491(4)  & 0.4305(1)  & 13.923(7) & 2.940(1)   & 0.1599(2) \\
\hline
$\infty$  & 5.5007(14)& 0.4302(9)& 13.92(6) & 2.934  & 0.15991(5) \\
\hline
\hline
\end{tabular}
\end{center}
\end{table}

To verify that the leading scaling corrections are absent in any 
observable for $d/b\approx 0.275$, we consider other observables. 
First, we check that the radius of gyration scales as $L^{\nu_{GS}}$
with tiny corrections.
If we use the numerical result $R_g = 20.882(4)$ for $L_1 = 1000$ 
and \cite{Clisby-10} $\nu_{GS} = 0.587597$,
we predict for $L_2=100$
\begin{equation}
   R_g(L_2) = R_g(L_1) \left( {L_2\over L_1}\right)^{\nu_{GS}} \approx 
    5.397,
\end{equation}
which is in excellent agreement with the direct numerical estimate 
$R_g(L_2) = 5.3901(4)$. At the optimal value of $d/b$, the scaling 
$R_g \sim L^{\nu_{GS}}$ holds therefore with a 0.1\% error already for $L=100$.
As an additional check, we have measured some adimensional
quantities related to the polymer shape. If 
$\lambda_1 \le \lambda_2 \le \lambda_3$
are the eigenvalues of the gyration tensor, we consider
\cite{Sciutto-96,shape,Causo-02} the asphericity $A_{\rm sph}$
\begin{equation}
A_{\rm sph} = \left \langle {\lambda_1^2 + \lambda^2_2 + \lambda^2_3 - 
          \lambda_1 \lambda_2 - \lambda_1 \lambda_3 - \lambda_2 \lambda_3
   \over  (\lambda_1 + \lambda_2 + \lambda_3)^2 } \right\rangle
\end{equation}
and the ratios $L_{12} = \langle\lambda_2\rangle/\langle\lambda_1\rangle$,
$L_{13} = \langle\lambda_3\rangle/\langle\lambda_1\rangle$.
Finally, we also consider the ratio $A_{ge} = R^2_g/R_e^2$, where 
$R^2_e$ is average squared end-to-end distance.
Results are reported in Table~\ref{tab:dstar}. The $L$-dependence is tiny
and significantly smaller than that observed for lattice self-avoiding
walks (SAWs) \cite{Sciutto-96,shape,Causo-02}, 
for which scaling corrections are large. Extrapolations
give results that are consistent with the extrapolated SAW
results. As a final check we have also computed the 
effective two-body and three-body center-of-mass effective potentials
\cite{BLHM-01,Louis}, see Figs.~\ref{fig:Vstar} and \ref{fig:V3star}.
The results for the two-body potentials can be compared with 
those obtained for SAWs (extrapolations of results for chains 
of length up to $L=4000$) \cite{BLHM-01,PH-05}, and with the 
ones \cite{DPP-compressible} obtained by using the 
lattice Domb-Joyce model \cite{DJ-72}. 
We observe perfect agreement, already for chains of 
$L=250$ monomers. The same holds for the three-body potential on equilateral
configurations.

\section{Comparison with other popular continuum models} \label{sec5}

\begin{figure}[tb!]
\begin{center}
\begin{tabular}{c}
\includegraphics[width=0.45\textwidth,angle=0]{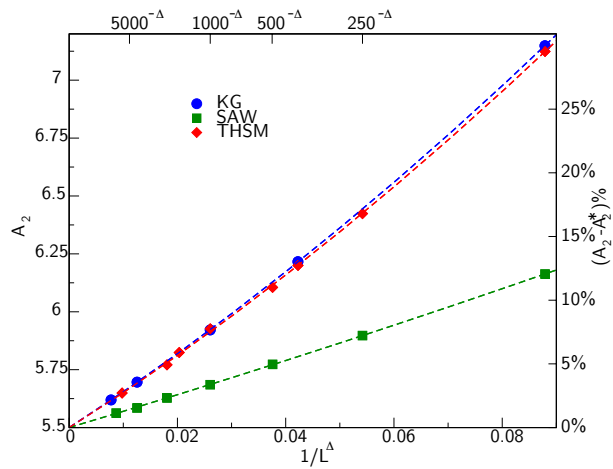} \\
\end{tabular}
\end{center}
\caption{Second-virial combination $A_2$ vs $L^{-\Delta}$, $\Delta = 0.528$,
for the KG model with $\beta\epsilon_{nb} = 4$ (KG) and 
for the THSM with $d=b$ (THSM). For comparison, we also report results for 
the self-avoiding walk (SAW) model on a cubic lattice. 
The asymptotic estimate is $A_2^* = 5.5007$ \cite{DP-16}. 
On the left we report the relative deviation $(A_2-A_2^*)/A^*_2$ in \%. }
\label{fig:A2}
\end{figure}

\begin{figure}[tb!]
\begin{center}
\begin{tabular}{c}
\includegraphics[width=0.45\textwidth]{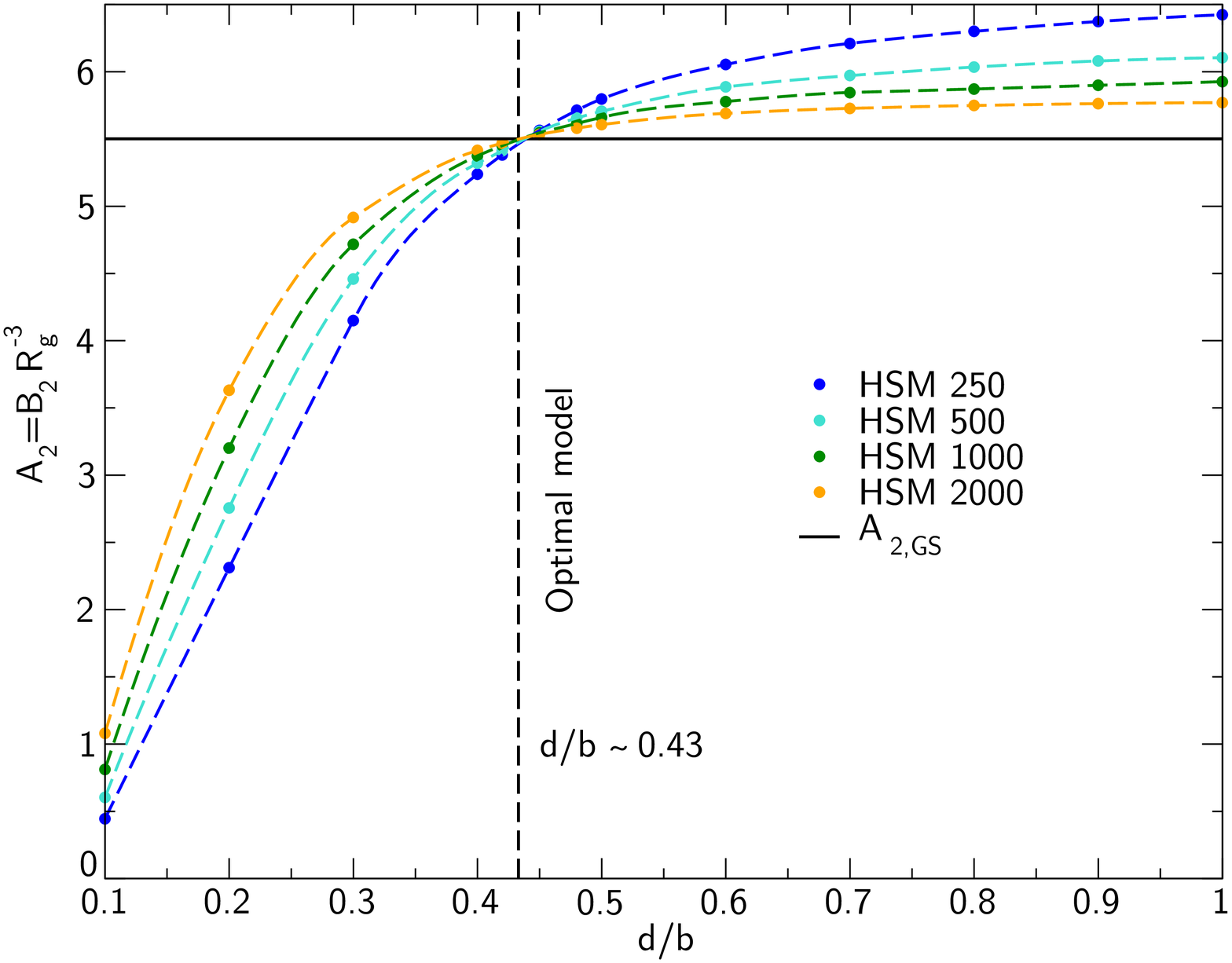} \\
\includegraphics[width=0.45\textwidth]{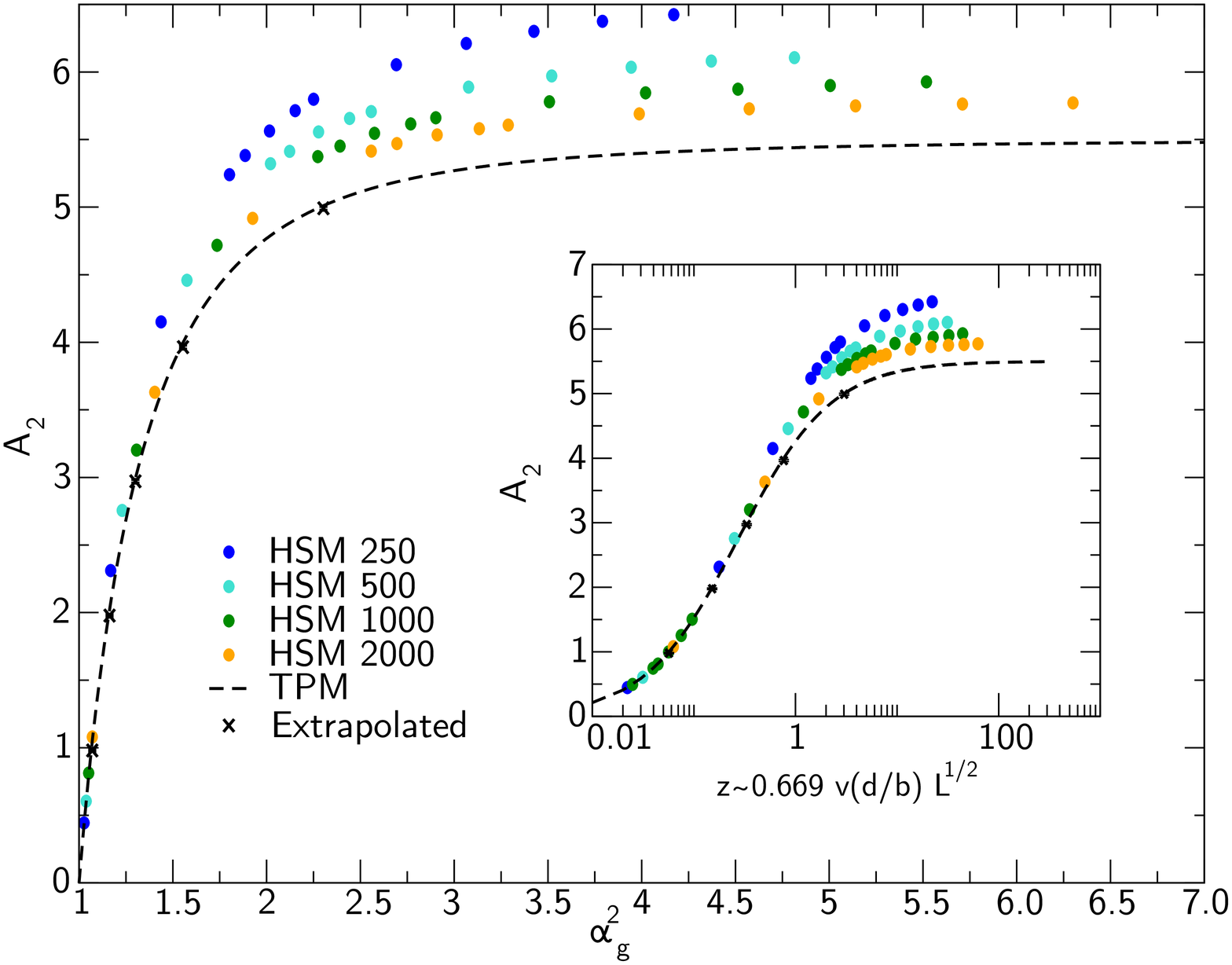} \\
\end{tabular}
\end{center}
\caption{Crossover behavior of the generalized hard-sphere model. Top: 
$A_2$ versus $d/b$ for several values of $L$. Bottom: $A_2$ versus $\alpha^2_g$
and (inset) $A_2$ versus $z$. The same data in the spherocylinder case are 
shown in Figs.~\ref{fig:A2vsd} and \ref{fig:A2-cross}.
}
\label{fig:GHSM-crossover}
\end{figure}

In the analysis of the behavior of polymer solutions, two other models 
are commonly used, the tangent-hard
sphere model (THSM) \cite{DH-86,DH-88}, and the Kremer-Grest (KG) model
\cite{GK-86}. As we shall show, these models exhibit very large finite-length
corrections, so values of $L$ of the order of $10^3$ are required to obtain
results that differ from the asymptotic ones by less than 10\%.
Both models can be generalized.
By tuning a single parameter that plays the same role as the thickness $d$,
one can define optimal models that show GS behavior for small values of $L$. 
Moreover, also these generalized models can be used to describe the 
thermal crossover.
However, because different bonds can cross each other, they
cannot be easily employed to study the polymer dynamics or 
topological properties.

\subsection{Definition of the generalized models} \label{sec5.1}

The generalization of the THSM is defined as follows. A chain of $L$ 
monomers is a random walk $\{{\bm r}_1,\ldots,{\bm r}_L\}$, such that 
$|{\bm r}_i - {\bm r}_{i+1}| = b$ ($b$ is therefore the bond length)
for all successive monomers. Each monomer is modelled as a hard sphere of 
diameter $d$, so  that $|{\bm r}_i - {\bm r}_{j}| > d$ if $|i-j|>1$. The
THSM is obtained by setting $d = b$. In the generalized model, 
by decreasing $b/d$, one can increase the stiffness of the polymers.
The angle $\phi_i$ defined in (\ref{def-phi}) is always smaller
than $\phi_0$, with $\phi_0 \to 180^\circ$ for $b/d\to \infty$ (completely
flexible chains),
$\phi_0 = 120^\circ$ for $b/d=1$ and $\phi_0\to 0$ as $b/d\to1/2$ (rod limit).
Because of this property, such a model has already been used 
to model semiflexible protein chains \cite{SHG-16}.

A generalization of the KG model is 
obtained by considering different Lennard-Jones potentials for the bonding and
non-bonding interactions. To be precise, we define the truncated and shifted 
Lennard-Jones potential
\begin{equation}
V_{LJ}(r,\epsilon) = \begin{cases} \epsilon  \left[
   \left({\sigma\over r}\right)^{12} - \left({\sigma\over r}\right)^{6} + 
   {1\over 4} \right] & \qquad \mbox{for } {r\over\sigma} < 2^{1/6}, \\
    0 & \qquad \mbox{for } {r\over\sigma} \ge 2^{1/6}
\end{cases}
\end{equation}
and 
\begin{equation}
V_{\rm FENE}(r,\epsilon_F) = \begin{cases}
  - \epsilon_F \left( {b_0\over \sigma}\right)^2 
    \ln \left(1 - {r^2\over b_0^2}\right), 
    & \qquad r < b_0 \\
   +\infty  & \qquad r \ge b_0.
\end{cases}
\end{equation}
In the generalization of the KG model, bonded monomers interact with 
potential
\begin{equation}
V_{\rm bond}(r) = V_{LJ}(r,\epsilon_b) + V_{\rm FENE}(r,\epsilon_F),
\end{equation}
where $r = |{\bm r}_i - {\bm r}_{i+1}|$.
Parameters $\epsilon_b$, $\epsilon_F$ and $b_0$ are chosen as in 
\cite{GK-86}: $\beta \epsilon_b = 4$,
$\beta \epsilon_F = 15$, $b_0/\sigma = 1.5$. This choice guarantees that the 
typical bond length $b$ is approximately $\sigma$ (simulations indicate that 
$b\approx 0.97\sigma$ for an isolated chain for all different nonbonded
interactions we have considered). For nonbonded monomers 
we consider instead 
\begin{equation}
V_{\rm nonbond}(r) = V_{LJ}(r,\epsilon_{nb}),
\end{equation}
with the same $\sigma$, but with a different $\epsilon_{nb}$. The usual 
KG interaction is obtained for $\epsilon_{nb} = \epsilon_b$. 

The THSM and the standard KG models show significant finite-length
corrections, as it can be seen in Fig.~\ref{fig:A2}, where we report $A_2$ as 
a function of $L$. The data clearly approach the asymptotic value
\cite{CMP-06,DP-16}
$A_{2,GS}\approx 5.50$.
More quantitatively, if we fit the KG data at $\beta \epsilon_{nb} = 4$ to
$a + b/L^{\Delta} + c/L$, with $\Delta = 0.528$, we obtain 
$a = 5.498(15)$, in perfect agreement with the lattice result. 
However, in these models $A_2$ approaches the limiting value quite slowly:
the relative difference $A_2/A_{2,GS} - 1$  is
less than 10\% only for $L\gtrsim 10^3$. 
Similar results are obtained for the THSM \cite{DP-16}.
Note that, for a given value of $L$,
the two models give very close estimates of $A_2$. This can be rationalized 
by using the Barker-Henderson \cite{BH-67} 
mapping of the KG model onto the THSM.
In this approach the KG model is equivalent to the generalization of the 
THSM with $b \approx 0.97 \sigma$ and hard-sphere diameter
\begin{equation}
  d_{BH} = \int_0^\infty dr\, (1 - e^{-\beta V_{LJ}(r,\epsilon_{nb})}).
\end{equation}
For $\beta \epsilon_{nb} = 4$ we obtain $d_{BH} = 1.01\sigma\approx 1.04 b$.
The bond length and the hard-sphere diameter are approximately the same, 
so the KG model is essentially equivalenty to the THSM. 

As $A_{2}(L)$ in both models is larger than $A_{2,GS}$, an optimal model 
is obtained by reducing the strength of the nonbonding potential, i.e.,
by increasing $b/d$ or decreasing $\beta \epsilon_{nb}$. This analysis will
be presented below.

\subsection{Generalized hard-sphere model} \label{sec5.2}

\begin{center}
\begin{table*}[tb!]
\caption{Crossover behavior in the generalized hard-sphere model.
We perform runs at
values of $d/b$ such that $A_{2,TPM}(z) = 1,2,3,4,5$, where
$A_{2,TPM}(z)$ is given in Eq.~(\ref{A2TPM}) and
$z = 0.669 v_{HS}(d) L^{1/2}$. We report the estimates of
$A_2$ and $\alpha_g^2$
for several values of $L$ and the corresponding large-$L$
extrapolation (we assume
\cite{DJ-72,CMP-08,BN-97}
$A_{2,L} = A_{2,\rm extr} + b_1 L^{-1/2} + b_2/L$).
Finally $\Delta = 100 (1 - B_{\rm Extr}/B_{TPM})$, $B = A_2$ or $\alpha_g^2$.
}
\label{tab:THSM-cross}
\begin{tabular}{c|ccccccccc}
\hline
\hline
\footnotesize
$z$ & $A_{2,TPM}(z)$ & 250 & 500 & 1000 & 2000 & 4000 & Extr. & $\Delta\%$ \\
\hline
0.056778 &
1 &     1.011(1) & 1.006(2) & 0.993(4) & 0.984(8) & 0.991(3) & 0.980(8) & 2\\
0.151493 &
2 &     2.091(2) & 2.055(3) &2.027(6) &2.02(1) & 2.003(5) &  1.98(1) &2 \\
0.331075 &
3 &  3.253(3) & 3.166(4) & 3.108(7) & 3.07(1) & 3.037(6) &  2.97(1) & 3 \\
0.767908 &
4 &  4.507(3) & 4.340(4) & 4.224(8) & 4.146(15) & 4.093(6) & 3.96(2) & 1 \\
3.01381 &
5 & 5.854(3) & 5.580(4) & 5.388(8) & 5.27(1) & 5.187(6) & 4.99(2) & 0.2 \\
\hline
\hline
& $\alpha^2_{g,TPM}(z)$ & 250 & 500 & 1000 & 2000 & 4000 & Extr. \\
\hline
0.056778 &
1.0668  & 1.0627(2)&  1.0638(3) & 1.0659(6) & 1.0668(13) & 1.0673(6) & 1.070(1)
& 0.32\\
0.151493 &
1.1602  & 1.1474(2) & 1.1516(3) & 1.1543(6)  &1.1591(13) & 1.1583(6) & 1.163(1)
&  0.2 \\
0.331075 &
1.3020  &  1.2785(2) & 1.2865(3) & 1.2916(7)  &1.2937(13) & 1.2967(6) &1.301(1)
& 0.1 \\
0.767908 &
1.5553  &  1.5249(2) &1.5371(4) & 1.5445(7)  &1.5481(13) & 1.5503(6) &1.555(1)
& 0.0\\
3.01381 &
2.2857  &  2.3201(3) & 2.3222(5) & 2.3233(8)  &2.319(2)  & 2.3136(7) & 2.303(5)
& 0.8\\
\hline
\hline
\end{tabular}
\end{table*}
\end{center}

We consider first the hard-sphere model, repeating the analysis 
presented in Sec.~\ref{sec4}. Simulations show different regimes,
see the top panel of 
Fig.~\ref{fig:GHSM-crossover}, which are completely analogous 
to those observed for spherocylinders. For $d/b\lesssim 0.43$,
there is a clear crossover between the ideal and the GS behavior.  
For $d/b\approx 0.43$, the $L$ dependence is tiny, while, for $d/b\gtrsim 0.43$,
$A_2$ is larger than $A_{2,GS}$ for finite values of $L$. The crossover 
behavior for $d/b\lesssim 0.43$ can be parametrized by using the variable 
$z$, which is now defined as $z = a_{HS} v_{HS}(d/b) L^{1/2}$, where 
$v_{HS}(d/b) = {2\pi\over 3} d^3/b^3$ is the adimensional 
monomer-monomer second-virial
coefficient. The constant $a_{HS}$ is again fitted so as to reproduce the data 
for small values of $z$. Using data obtained from simulations
of chains with $L=1000$, we obtain $a_{HS} = 0.669(2)$. As 
in the spherocylinder case, the TPM curves describe quite well the 
crossover up to $A_2\approx 5$, see Fig.~\ref{fig:GHSM-crossover}, while 
larger discrepancies are observed for larger values. We also
perform a detailed check for a few selected values of $z$. The results, shown
in Table \ref{tab:THSM-cross}, confirm the identification of the $z$ variable.
In all cases, the extrapolated (large-$L$) results are consistent with 
the TPM predictions.

\begin{figure}[h!]
\begin{center}
\begin{tabular}{c}
\includegraphics[width=0.45\textwidth]{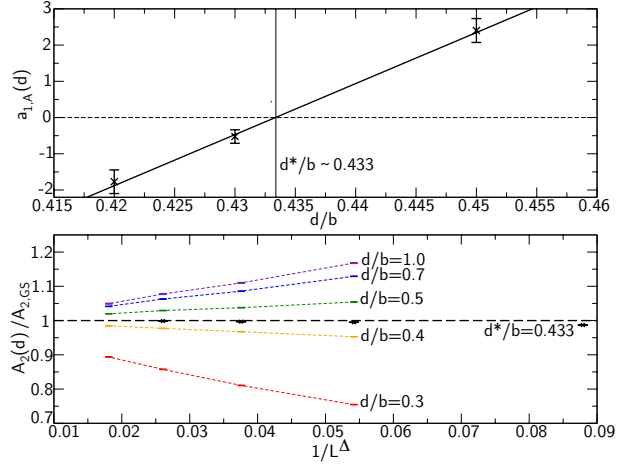} \\
\end{tabular}
\end{center}
\caption{Top: Scaling correction coefficient $a_1(d)$ versus $d/b$ for the
generalized hard-sphere model; the line is the linear interpolation.
Bottom: $A_2(d,L)/A_{2,GS}$ versus $L^{-\Delta}$, $\Delta = 0.528$,
$A_{2,GS} = 5.500$,
for several values of $d/b$.}
\label{fig:A2-LmDelta-HSM}
\end{figure}

We also determined the optimal model for which there are no
leading finite-length corrections. For this purpose,
we performed simulations of chains of length 250, 500, 1000, and 
2000 for $d/b = 0.42,0.43,0.45$. Fits to Eq.~(\ref{fitA2}) give estimates
of $a_1(d/b)$. Performing a linear interpolation, see 
Fig.~\ref{fig:A2-LmDelta-HSM}, we obtain 
\begin{equation}
d^*/b = 0.433(1).
\end{equation}
This result compares well with the estimate 
$d^*/b \approx 0.447$ of \cite{LK-99}. 

\begin{table}[tb!]
\caption{Second-virial combination $A_2$, asphericity $A_{\rm sph}$, 
ratios $L_{3,1}$ and $L_{2,1}$ 
between the eigenvalues of the gyration tensor, and ratio $A_{ge}$, 
for the optimal 
hard-sphere model with $d/b = 0.433$ and several values of $L$. 
The extrapolated value of $A_2$ is taken from \cite{DP-16},
those for the shape factors are extrapolations of the
results of \cite{Sciutto-96}, while the
estimate of $A_{ge}$ is taken from \cite{Clisby-10}.
}
\label{tab:optimalTHSM}
\begin{tabular}{c|ccccc}
\hline
\hline
\footnotesize
$L$       & $A_{2}$  & $A_{\rm sph}$   &  $L_{3,1}$   & $L_{2,1}$  &
   $A_{ge}$ \\
\hline
100       & 5.430(3)  & 0.42757(5) &  13.691(4) & 2.9061(7) & 0.16137(5) \\
250       & 5.475(4)  & 0.42930(7) &  13.824(5) & 2.9249(9) & 0.16049(7) \\
500       & 5.478(4)  & 0.42994(9) &  13.875(6) & 2.9321(12)& 0.16024(8)\\
1000      & 5.509(8)  & 0.4301(2)  &  13.890(11)& 2.935(2)  & 0.1601(1)\\
\hline
$\infty$  & 5.5007(14) & 0.4302(9) &  13.92(6)  & 2.934  & 0.15991(5) \\
\hline
\hline
\end{tabular}
\end{table}
As we have already mentioned, in the optimal model {\em any} observable should
not present scaling corrections decaying as $L^{-\Delta}$, guaranteeing 
a faster convergence to the infinite-length limit. 
To verify this cancellation,  we consider 
again the shape factors. Results for the 
hard-sphere model are reported in 
Table~\ref{tab:optimalTHSM}. They should be compared with 
those appearing in Table~\ref{tab:dstar} for the  
spherocylinder model. Also in the hard-sphere case do we observe 
a fast convergence (discrepancies are well explained by corrections
decaying as $1/L$). As for the next-to-leading corrections, the 
optimal spherocylinder model performs better than the hard-sphere one.

\begin{center}
\begin{table*}
\caption{Estimates of $A_2$ as a function of the non-bonding LJ parameter
$\beta \epsilon_{nb}$ and of $L$ for the generalized Kremer-Grest model.
Here $d_{BH}$ is the Barker-Henderson effective diameter [for
small values of $\epsilon_{nb}$, one can use the asymptotic
expression $d_{BH}/\sigma \approx (\beta \epsilon_{nb})^{1/12} \Gamma(11/12)
\approx 1.056(\beta \epsilon_{nb})^{1/12} $].
The column $L=\infty$ gives the estimates of $A_{2,GS}$ obtained by
fitting the data to $A_{2,GS} + a/L^\Delta + b/L^{\Delta_2}$,
fixing $\Delta = 0.528$ and $\Delta_2 = 1$ (errors, in parentheses,
take into account the uncertainty on these two exponents).
}
\label{table:GK}
\begin{tabular}{cccccccc}
\hline\hline
\footnotesize
$\beta \epsilon_{\rm nb}$ & $d_{BH}/\sigma$ &
             $L = 100$ & $L=400$ & $L=1000$ & $L=4000$ & $L=10000$  &
$L=\infty$ \\
\hline
4.0 & 1.01  & 7.149(3) &  6.216(3) & 5.921(3) & 5.696(8)  & 5.619(10)
                   & 5.498(15) \\
0.001&0.59  & 6.470(2) &  5.963(2) & 5.781(4) & 5.633(6) & & 5.491(17) \\
0.0001&0.49 & 6.025(3) &  5.774(2) & 5.673(4) & 5.579(7) & & 5.491(17) \\
0.00002&0.43& 5.566(1) &  5.556(2) & 5.541(2) & 5.523(4) & & 5.502(9) \\
0.00001&0.40& 5.329(1) &  5.433(2) & 5.459(3) & 5.484(5) & & 5.497(9)\\
\hline\hline
\end{tabular}
\end{table*}
\end{center}

\begin{figure}[t!]
\begin{center}
\begin{tabular}{c}
\includegraphics[width=0.45\textwidth]{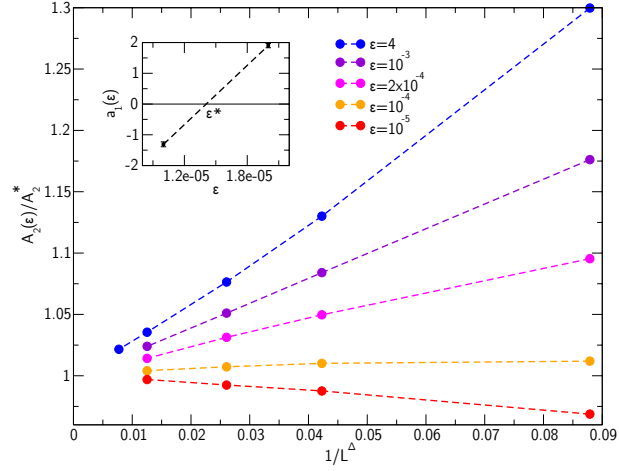} \\
\end{tabular}
\end{center}
\caption{
$A_2(d,L)/A_{2}^*$ ($A_2^* = A_{2,GS} = 5.500$)
versus $L^{-\Delta}$, $\Delta = 0.528$,
for several values of $\beta \epsilon$ in the generalized KG model.
Inset: Scaling correction coefficient $a_1(d)$ versus $\beta \epsilon$;
the line is the linear interpolation.
}
\label{fig:A2-LmDelta-GK}
\end{figure}

\subsection{Generalized Kremer-Grest model} \label{sec5.3}

The same analysis can be performed in the KG case. We first determine 
the optimal model, considering several 
values of $\epsilon_{nb}$ in the range 
$10^{-5} \le \beta \epsilon_{nb} \le 10^{-3}$ and values of $L$ in the 
range $100\le L \le 4000$. The results are reported in 
Table~\ref{table:GK} and shown in Fig.~\ref{fig:A2-LmDelta-GK}. 
In the limit $L\to \infty$ all data converge to the 
same value, in agreement with universality. However, for finite 
$L$, deviations are typically 
large, except when $\beta\epsilon_{nb}$ is approximately $10^{-5}$.
Repeating the analysis presented for the spherocylinder model,
we obtain the optimal value,
see Fig.~\ref{fig:A2-LmDelta-GK},
\begin{equation}
\beta \epsilon_{nb}^* = 1.41(6) \cdot 10^{-5},
\end{equation}
where the error (in parentheses) 
takes also into account the uncertainties on $A_{2,GS}$,
$\Delta$, and $\Delta_2$. It is interesting to compare this result with 
the one for the hard-sphere model using the Barker-Henderson mapping. 
As $\beta \epsilon_{nb}^*$ is very small, we can use the asymptotic behavior
$d_{BH}/\sigma \approx (\beta \epsilon_{nb})^{1/12} \Gamma(11/12) \approx
1.056 (\beta \epsilon_{nb})^{1/12}$  (see Appendix for the derivation).
For the optimal model we obtain $d_{BH}\approx 0.416\sigma  
\approx 0.43 b$, which is the value obtained above for the hard-sphere model.

The behavior of the generalized KG model for $\epsilon_{nb} < \epsilon_{nb}^*$
is again expected to be described by the TPM provided we define 
$z$ as $z = a_{KG} v(\epsilon_{nb}) L^{1/2}$ with 
\begin{equation}
   v(\epsilon_{nb}) = {1\over 2\sigma^3} \int d^3r\, 
(1 - e^{-\beta V_{LJ}(r,\epsilon_{nb})}).
\end{equation}
Since we will be interested in values of $\epsilon_{nb}$ smaller than 
$\epsilon^*_{nb}$, we can replace the previous expression with its 
asymptotic behavior (see Appendix), $v(\epsilon_{nb}) \approx (2 \pi/3) (\beta
\epsilon_{nb})^{1/4} \Gamma(3/4)
\approx 2.567 (\beta \epsilon_{nb})^{1/4}$. 
To compute $a_{KG}$, we take advantage of the Barker-Henderson mapping. 
Since we expect $z$ to be approximately equal to $a_{HS} v_{HS}(d_{BH}/b)
L^{1/2}$, comparing the two expressions and using $b/\sigma \approx 0.97$,
we obtain
$a_{KG} \approx 0.70$. 
Therefore, we can simply define
\begin{equation}
z = 1.80 (\beta \epsilon_{nb})^{1/4} L^{1/2}.
\end{equation}
To verify this prediction, we have 
performed simulations for $\beta\epsilon_{nb} = 8\cdot 10^{-13}$. 
Using the TPM expression (\ref{A2TPM}) we predict
$A_2\approx 0.96$ for this value of the nonbonding parameter.
Simulations for $L=1000$ give $A_2 = 0.992(3)$, in reasonable agreement.

\section{Conclusions} \label{sec6}

One of the basic tenets of the theory of polymer solutions is the 
concept of universality \cite{deGennes-79,Freed-87,dCJ-book,Schaefer-99}.
The thermodynamic behavior and large-scale structure 
of these typical soft-matter systems
can be described in a wide concentration range (the so-called dilute and 
semi-dilute regimes) by using a limited number of microscopic variables
that depend on the specificities of the experimental system. Because of 
universality, theoretical predictions can be obtained by using simplified 
models, in which only a few of the properties of the original system are 
retained. Lattice self-avoiding walks have been extensively used and 
so have several continuum models, like the THSM and the KG model
\cite{DH-86,DH-88,GK-86}. Although they all have the same behavior 
for long polymers, for chains of a few hundred monomers (this is the typical
number in finite-density simulations) they give predictions that significantly
differ from the asymptotic value. For instance, for $L=100$ the THSM and the 
KG model overestimate $A_2$ (or the equivalent interpenetration ratio 
$\Psi$) by 30\%, while the self-avoiding walk is only slightly more 
accurate, with a discrepancy of 12\%. Even worse, discrepancies increase 
as one enters the semidilute regime. For instance, if $Z = \beta P/\rho_p$ 
($P$ is the pressure and $\rho_p$ the polymer number density) is the 
compressibility factor, \cite{JHGJM-12} reports 
$Z \approx 100$ for a polymer packing fraction
$\phi_p=10$ if one uses the THSM with $L= 100$. This is a factor of three 
larger than the result $Z \approx 34$ obtained by using an equation of state 
appropriate for very long good-solvent polymers \cite{Pelissetto-08}.
It is therefore clear that in the semidilute region the THSM with 
such a number of monomers is far from 
the universality regime, and hence its predictions do not agree with 
what would be obtained in a different model or in an experimental system. 

To overcome these problems, one can consider optimal models in which 
the approach to the universal limit is faster. The Domb-Joyce model \cite{DJ-72}
is a simple generalization of the more common self-avoiding walk model, 
that allows one to obtain asymptotic results for small monomer numbers
by appropriately choosing the value of the energy penalty
\cite{BN-97,CMP-06,DP-16,Clisby-17}. There are, however, situations, 
in which a lattice model or a model in which self-intersections are allowed is 
not convenient. For instance, one cannot study solid phases (this is of 
interest for high-functionality star polymers \cite{star}) 
nor use it in all those 
contexts in which concatenation and topology are important. 

In this paper we consider two continuum polymers models,
which are simple generalizations of the THSM and of the 
KG model. By an appropriate choice of a single parameter, 
we obtain improved models which show a fast 
convergence. They can be conveniently used to study 
thermodynamics and phase diagrams, but, since in both cases bonds can cross,
they are not appropriate for dynamical or topological studies.
Thick self-avoiding chains, obtained by linking spherocylinders of length
$b$ and diameter $d$, have none of these limitations. Bond intersections 
are strictly forbidden and there exists a value of the ratio $d/b$ 
($d/b \approx 0.275$) for which the model is optimal. For $d/b > 0.275$
corrections increase with $d/b$, so the largest $d/b$ is, the largest 
the number $L$ of monomers has to be in order to observe the universal 
asymptotic behavior. In the opposite regime we observe a crossover 
between ideal ($d/b \approx 0$) and good-solvent behavior, which can be
described by using the two-parameter model
\cite{Yamakawa-71,dCJ-book,Schaefer-99,CMP-08}, which parametrizes the 
temperature dependence in the vicinity of the $\theta$ point.  

It is important to stress that, although we have determined the 
optimal interaction by considering linear chains,
the optimal models show a fast convergence to the asymptotic limit 
for any polymer topology. For instance, the optimal Domb-Joyce model determined
by considering linear chains \cite{CMP-06,DP-16,Clisby-17} is also optimal when 
applied to star polymers, see the extensive analysis presented in
\cite{RP-13}. Therefore, the results apply to any type of highly
functionalized branched polymers. It is also interesting to investigate the 
behavior of cyclic chains. Renormalization-group arguments guarantee 
the optimality of
the models as long as no constraint on the polymer topology is imposed, i.e., 
if one averages over all cyclic chains, independently of the knot type. 
If one considers instead 
polymers of fixed knot type, the behavior is less clear, as new 
corrections might appear. It must be noted, however, that all numerical 
simulations, see, e.g., \cite{BO-12}, 
indicate the presence of finite-size corrections decaying as 
$L^{-\Delta_k}$, with $\Delta_k \approx 0.5$ for all knot types. These 
results suggest 
that $\Delta_k$ coincides with the standard exponent $\Delta$, 
so that the corrections observed at fixed knot type are associated
with the renormalization-group operator that controls the approach 
to the universal limit in any polymer model. If this is the case, 
the model with optimal interaction should 
be optimal also for chains of fixed knot type. 
This issue is presently under investigation.

Finally, it is tempting to use our results for 
thick self-avoiding chains to get some physical insight on the 
behavior of double-stranded 
DNA filaments. In particular, we can 
address the question under which conditions DNA can be considered as 
a model good-solvent polymer. The ratio $d/b$ can be effectively changed by 
varying the ionic strength $I$ of the solution.
Using the results of \cite{Stigter-77,RCV-93,VC-95} and 
assuming that the effective 
diameter $d$ scales as $1/\sqrt{I}$, in analogy with the behavior of 
 the Debye length 
that sets the scale for the electrostatic interactions, we obtain 
$d = d_0/\sqrt{I}$, where $d_0 = 1.6$-1.9 nm$\cdot$M$^{1/2}$ for a monovalent
salt (Na$^+$ for instance). For the Kuhn length $\ell_K$ 
several results are available
in the literature,\footnote{
We only consider a monovalent salt.
For the persistence length $\ell_p = \ell_K/2$
we use two different interpolating formulas:
(a) $\ell_p({\rm nm}) = 44 + 
0.0136/I({\rm M})$ (reference \cite{DC-09});
(b) $\ell_p({\rm nm}) = 33.8 + 24.3/(1 + 8.21 I^{0.93})$, where $I$ is measured 
in M = mol/L \cite{BTSRDM-15}.
Other results are cited in \cite{DC-09,BTSRDM-15}.
Formula (a) predicts $\ell_K = 115, 88$ nm for $I = 0.001, 0.2$ M, 
respectively;
formula (b) predicts $\ell_K = 116, 85$ nm for the same values of $I$.
}
which all predict $\ell_K$ of the 
order of 100 nm for $I$ varying between 1 mM and 200 mM.
We assume that our model provides a realistic 
description of DNA provided that $b$ is identified with $\ell_K$. 
Therefore, we predict that good-solvent
behavior for relatively small values of $L$ should be observed 
for $d/\ell_K \approx 0.27$.
Using the results reported above, we can correspondingly obtain 
the optimal ionic strengh: $I = 3$-6 mM
for a monovalent salt. The corresponding Kuhn length 
is $\ell_K = 90$-110 nm, i.e., approximately 300 base pairs. Therefore,
for values of $I$ in the optimal range, 
one should be able to observe good-solvent 
behavior by using DNA of 30000 base pairs.

The optimal value of $I$ is, however,
well below physiological conditions ($I\sim 100$-200 mM).  
If we take $I = 150$ mM, we predict $d/\ell_K\approx 0.05$.
Also in this regime we can predict theoretically the observed
behavior, as long as the DNA filaments are significantly longer than
the Kuhn length. We can indeed use the
wealth of results available for the
TPM \cite{Yamakawa-71,Schaefer-99,dCJ-book,CMP-08,DPP-13-Thermal}.
The only external input is the model-dependent constant $a$ entering the 
definition of the TPM variable $z$. If we assume that our model provides 
a reasonable description of DNA, we predict $z\approx z_0 L^{1/2}$, 
where $z_0 \approx 0.03$-0.04, and $L$ is the ratio between 
the length of the filament and the Kuhn length. 
Note that, for such small values of $d/b$, 
scaling corrections are large (see Fig.~\ref{fig:A2vsd} for instance)
and good-solvent behavior may be  unattainable in practice,
in agreement with the conclusions of \cite{TMDD-13}.
Indeed, as good-solvent behavior is (very roughly)
observed for $z\gtrsim 5$ (see the supplementary material of 
\cite{DPP-14-GFVT}), DNA is close to the universal asymptotic regime
only for $L\gtrsim 20000$-30000, i.e.,  when the filament has 
at least $10^6$ base pairs.

\bigskip
G.D. thanks C. Micheletti and C. Pierleoni for fruitful discussions.

\appendix

\section{Computation of integrals for the generalized Kremer-Grest model}

In Section \ref{sec5.3} 
we use the asymptotic expansion of some integrals involving
the Kremer-Grest potential. The general integral we consider is 
\begin{equation}
I(s,\alpha) = \int_0^a dx\ x^s \left\{1 - 
    \exp\left[-\alpha \left({1\over x^{12}} - {1\over x^6} + 
    b\right)\right]\right\},
\end{equation}
where $-1 <  s < 5$, $a$ is a positive finite number and $b$ is a second
constant. We wish to compute the 
expansion of this integral for $\alpha \to 0$. First, we rewrite the integral
as
\begin{eqnarray}
I(s,\alpha) &=& (1 - e^{-b \alpha}) \int_0^a dx\ x^s 
\nonumber \\
     &-& e^{-b \alpha} \int_a^\infty dx\ x^s 
     \left(1 - e^{-\alpha/x^{12} + \alpha/x^6}\right) 
\nonumber \\
     &+& e^{-b \alpha} \int_0^\infty dx\ x^s 
     \left(1 - e^{-\alpha/x^{12} + \alpha/x^6}\right). 
\end{eqnarray}
The first term is obviously of order $\alpha$. In the second term, 
for $s < 5$ we can expand the exponential to first order in $\alpha$, 
proving that this contribution is of order $\alpha$, too. Therefore, we obtain  
\begin{equation}
I(s,\alpha) = 
     \int_0^\infty dx\ x^s 
     \left(1 - e^{-\alpha/x^{12} + \alpha/x^6}\right) + O (\alpha). 
\end{equation}
Now, we change variable, defining $y = \alpha/x^{12}$. We obtain 
\begin{eqnarray}
&& I(s,\alpha) \approx 
   {\alpha^{(s+1)/12} \over 12} 
   \int_0^\infty dy\ y^{-(s+13)/12} (1 - e^{-y + \sqrt{\alpha y}}) 
\nonumber \\
&& \quad  \approx  {\alpha^{(s+1)/12} \over 12} 
   \int_0^\infty dy\ y^{-(s+13)/12} (1 - e^{-y}) 
\nonumber \\
&& \quad = 
   -{\alpha^{(s+1)/12} \over 12} \Gamma\left(- {1+s\over12}\right) 
\nonumber \\ 
&& \quad = 
   {\alpha^{(s+1)/12} \over s+1} \Gamma\left({11-s\over12}\right) 
+ O(\alpha^{(s+7)/12}),
\end{eqnarray}
where, in the last line, we have explicitly reported the order of the neglected
terms.


\section*{References}

\begin{thebibliography}{99}

\bibitem{Binder-96}
Binder K (ed.) 1996
{\em Monte Carlo and Molecular Dynamics Simulations in Polymer
Science} (Oxford, UK: Oxford University Press)

\bibitem{deGennes-79}
de Gennes P G 1979
{\em Scaling Concepts in Polymer Physics} 
(Ithaca, NY: Cornell University Press)

\bibitem{Freed-87}
Freed K F 1987
{\em Renormalization Group Theory of Macromolecules}
(New York: Wiley)

\bibitem{DE-88}
Doi M and Edwards S F 1988
{\em The Theory of Polymer Dynamics}
(Oxford: Clarendon Press)

\bibitem{dCJ-book}
des Cloizeaux J and Jannink G  1990
{\em Polymers in Solution: Their Modelling and Structure}
(Oxford: Clarendon Press)

\bibitem{Schaefer-99}
Sch\"afer L 1999
{\em Excluded Volume Effects in Polymer Solutions} 
(Berlin: Springer).

\bibitem{VFDLRRD-05}
Valle F, Favre M, De Los Rios P, Rosa A and Dietler G 2005
Scaling exponents and probability distributions of DNA 
end-to-end distance
{\em Phys. Rev. Lett.} {\bf 95} 158105

\bibitem{NYSK-13}
Nepal M, Yaniv A, Shafran E and Krichevsky O 2013
Structure of DNA coils in dilute and semidilute solutions,
{\em Phys. Rev. Lett.} {\bf 110} 058102

\bibitem{TMDD-13}
Tree D R, Muralidhar A, Doyle P S and Dorfman K D 2013
Is DNA a good model polymer?
{\em Macromolecules} {\bf 46} 8369

\bibitem{DH-86}
Dickman R and Hall C K 1986
Equation of state for chain molecules:
Continuous-space analog of Flory theory
{\em J. Chem. Phys.} {\bf 85} 4108

\bibitem{DH-88}
Dickman R and Hall C K 1988
High-density Monte Carlo simulations of chain molecules: Bulk equation
of state and density profile near walls
{\em J. Chem. Phys.} {\bf 89} 3168

\bibitem{GK-86}
Grest G S and Kremer K 1986
Molecular dynamics simulation for polymers in the presence of a heat bath
{\em Phys. Rev.} A {\bf 33} 3628

\bibitem{deGennes-72}
de Gennes P G 1972
Exponents for the excluded volume problem as derived by the Wilson method
{\em Phys. Lett.} A {\bf 38} 339

\bibitem{Wegner-76}
Wegner F J 1976
The Critical State, General Aspects
{\em Phase Transitions and Critical Phenomena}
Domb C and Green M S (eds) vol 6
(New York: Academic Press)

\bibitem{PHA-91}
Privman V, Hohenberg P C and Aharony A 1991
Universal Critical-Point Amplitude Relations
{\em Phase Transitions and Critical Phenomena}
Domb C and Lebowitz J (eds) vol 14
(London: Academic Press)

\bibitem{PV-02}
Pelissetto A and Vicari E 2002
Critical phenomena and renormalization-group theory
{\em Phys. Rep.} {\bf 368} 549

\bibitem{Oono-85} 
Oono Y 1985
Statistical physics of polymer solutions: Conformation-space
renormalization-group approach
{\em Adv. Chem. Phys.} {\bf 61} 301 (Sec. V)

\bibitem{WDF-86}
Wang S Q, Douglas J F and Freed K F 1986
Corrections to preaveraging approximation within the
Kirkwood-Riseman model for flexible polymers: Calculations to
second order in $\epsilon$ with both hydrodynamic and excluded
volume interactions
{\em J. Chem. Phys.} {\bf 85} 3674

\bibitem{WF-86}
Wang S Q and Freed K F 1986
Renormalization group study of Rouse-Zimm model of polymer
dynamics through second order in $\epsilon$
{\em J. Chem. Phys.} {\bf 85} 6210

\bibitem{SFO-85}
Schaub B, Friedman B A and Oono Y 1985
Time-dependent correlations of a self-avoiding polymer chain
{\em Phys. Lett.} A {\bf 110} 136

\bibitem{SH-89}
Stepanow S and Helmis G 1989
Renormalization-group study of the dynamical
viscosity of dilute solutions of self-avoiding polymer chains
{\em Phys. Rev.} A {\bf 39} 6037

\bibitem{Grassberger-97}
Grassberger P 1997
Pruned-enriched Rosenbluth method: Simulations of theta polymers of chain
length up to 1,000,000
{\em Phys. Rev.} E {\bf 56} 3682

\bibitem{Clisby-10-JSP}
Clisby N 2010
Efficient implementation of the pivot algorithm for self-avoiding walks
{\em J. Stat. Phys.} {\bf 140} 349

\bibitem{BN-97}
Belohorec P and Nickel B G 1997
Accurate universal and two-parameter model results from a monte-carlo
renormalization group study
{\em Guelph University report} (unpublished) \par
Belohorec P 1997
Renormalization group calculation of the universal critical exponents of a
polymer molecule {\em Guelph University PhD thesis} (available at
{\tt \tiny www.collectionscanada.gc.ca/obj/s4/f2/dsk3/ftp04/nq24397.pdf}).

\bibitem{LK-99}
Lue L and Kiselev S B 1999
Crossover approach to scaling behavior in dilute polymer solutions: Theory and
simulation
{\em J. Chem. Phys.} {\bf 110} 2684 \par
Lue L and Kiselev S B 2001
Crossover behavior of star polymers in good solvents
{\em J. Chem. Phys.} {\bf 114} 5026

\bibitem{CMP-06}
Caracciolo S, Mognetti B M and Pelissetto A 2006
Virial coefficients and osmotic pressure in polymer solutions 
in good-solvent conditions,
{\em J. Chem. Phys.} {\bf 125} 094903

\bibitem{DP-16}
D'Adamo G and Pelissetto A 2016
Improved model for mixtures of polymers and hard spheres
{\em J. Phys.} A: {\em Math Theor.} {\bf 49} 504006

\bibitem{Clisby-17}
Clisby N 2017
High resolution Monte Carlo study of the Domb-Joyce model
{\tt arXiv:1705.01249 [cond-mat.stat-mech]}

% Accurate effective pair potentials for polymer solutions
\bibitem{BLHM-01}
Bolhuis P G, Louis A A, Hansen J P and Meijer E J 2001 
Accurate effective pair potentials for polymer solutions
{\em J. Chem. Phys.} {\bf 114} 4296

\bibitem{MullerPlathe-02}
M\"uller-Plathe F 2002
Coarse-graining in polymer simulation: From the atomistic to the mesoscopic
scale and back
{\em Chem. Phys. Chem.} {\bf 3} 754

\bibitem{PCH-07}
Pierleoni C, Capone B and Hansen J P 2007 
A soft effective segment representation of semidilute polymer solutions
{\em J. Chem. Phys.} {\bf 127} 171102

\bibitem{PK-09}
Peter C and Kremer K 2009 
Multiscale simulation of soft matter systems --- from the atomistic to the
coarse-grained level and back
{\em Soft Matter} {\bf 5} 4357

\bibitem{DPP-12-Soft}
D'Adamo G, Pelissetto A and Pierleoni C 2012 
Coarse-graining strategies for polymer solutions
{\em Soft Matter} {\bf 8} 5151

\bibitem{DPP-12-JCP}
D'Adamo G, Pelissetto A and Pierleoni C 2012 
Consistent and transferable coarse-grained model for semidilute polymer
solutions in good solvent
{\em J. Chem. Phys.} {\bf 137} 024901

\bibitem{KVMP-12}
Karimi-Varzaneh H A and M\"uller-Plathe F 2012 
Coarse-grained modeling for macromolecular chemistry
{\em Multiscale Molecular Methods in Applied Chemistry},
Kirchner B and Vrabec J (eds) 
{\em Top. Curr. Chem.} {\bf 307}
(Berlin:Springer) 295-321

\bibitem{Noid-13}
Noid W G 2013
Systematic methods for structurally consistent coarse-grained models,
{\em Methods Mol. Biol.} {\bf 924} 487\par
Noid W G 2013
Perspective: Coarse-grained models for biomolecular systems
{\em J. Chem. Phys.} {\bf 139} 090901

\bibitem{PB-01}
Padding J T and  Briels W J 2001
Uncrossability constraints in mesoscopic polymer melt simulations: Non-Rouse 
behavior of C$_{120}$H$_{242}$ 
{\em J. Chem. Phys.} {\bf 115} 2846

\bibitem{OW-07}
Orlandini E and Whittington S G 2007
Statistical topology of closed curves: Some applications in polymer physics
{\em Rev. Mod. Phys.} {\bf 79} 611-642

\bibitem{FSS-11}
Forgan R S, Sauvage J P and Stoddard J F 2011
Chemical Topology: Complex Molecular Knots, Links, and Entanglements
{\em Chem. Rev.} {\bf 111} 5434-5464

\bibitem{VRSWW-92}
Janse van Rensburg E J, Sumners D A W, Wassermann E and  
Whittington S G 1992
Entanglement complexity of self-avoiding walks
{\em J. Phys.} A: {\em Math. Gen.} {\bf 25} 6557-6566

\bibitem{TRFM-13}
Tubiana L, Rosa A, Fragiacomo F and Micheletti C 2013
Spontaneous Knotting and Unknotting of Flexible Linear Polymers: Equilibrium
and Kinetic Aspects
{\em Macromolecules} {\bf 46} 3669-3678

\bibitem{EB-14}
Evans N H and Beer P D  2014
Review Article: Progress in the synthesis and exploitation of catenanes since
the Millennium
{\em Chem. Soc. Rev.} {\bf 43} 4658-4683

\bibitem{WC-86}
Wassermann S A and Cozzarelli N R 1986 
Biochemical topology: Applications to DNA recombination and replication
{\em Science} {\bf 232} 951-960

\bibitem{VR-09}
Vologodskii A and Rybenkov V V 2009
Perspective: Simulation of DNA catenanes
{\em Phys. Chem. Chem Phys.} {\bf 11} 10543-10552

\bibitem{AA-80}
Aharony A and Ahlers G 1980
Universal ratios among correction-to-scaling amplitudes and effective 
critical exponents 
{\em Phys. Rev. Lett.} {\bf 44} 782

\bibitem{JRWM-90}
Janse van Rensburg E J, Whittington S G and Madras N 1990
The pivot algorithm and polygons: results on the FCC lattice
{\em J. Phys.} A: {\em Math. Gen.} {\bf 23} 1589

\bibitem{FZ-11}
Fuereder I and Zifferer G 2011
Monte Carlo simulation studies of ring polymers at athermal and theta
conditions
{\em J. Chem. Phys.} {\bf 135} 184906

\bibitem{HNG-04}
Hsu H-P, Nadler W and Grassberger P 2004
Scaling of star polymers with 1--80 arms
{\em Macromolecules} {\bf 37} 4658

\bibitem{RP-13}
Randisi F and Pelissetto A 2013
High-functionality star-branched macromolecules: Polymer size and virial 
coefficients
{\em J. Chem. Phys.} {\bf 139} 154902

\bibitem{Clisby-10}
Clisby N 2010
Accurate estimate of the critical exponent $\nu$ for self-avoiding walks
via a fast implementation of the pivot algorithm
{\em Phys. Rev. Lett.} {\bf 104} 055702

\bibitem{VLKFKC-92}
Vologodskii A V, Levene S D, Klenin K V, Frank-Kamenetskii M D and
Cozzarelli N R 1992
Conformational and Thermodynamic Properties of Supercoiled DNA
{\em J. Mol. Biol.} {\bf 227} 1224

\bibitem{RCV-93}
Rybenkov V V, Cozzarelli N R and Vologodskii A V 1993
Probability of DNA knotting and the effective diameter of the DNA double helix,
{\em Proc. Natl. Acad. Sci. (USA)} {\bf 90} 5307-5311

\bibitem{VC-95}
Vologodskii A and Cozzarelli N R 1995
Modeling of long-range electrostatic interactions in DNA
{\em Biopolymers} {\bf 35} 289-296

\bibitem{GM-99}
Gonzalez O and Maddocks J H 1999
Global curvature, thickness, and the ideal shapes of knots
{\em Proc. Natl. Acad. Sci.} (USA) {\bf 96} 4769 

\bibitem{SM-00}
Stasiak A and Maddocks J H 2000
Best packing in proteins and DNA
{\em Nature} {\bf 406} 251

\bibitem{MPR-08}
Millet K C, Piatek M and Rawdon E 2008
Polygonal knot space near ropelength-minimized knots
{\em J. Knot Theory Ramifications} {\bf 17} 601-631

\bibitem{UD-15-17}
Uehara E and Deguchi T 2015
Characteristic length of the knotting probability revisited
{\em J. Phys.: Condens. Matter} {\bf 27} 354104 \par
Uehara E and Deguchi T,
Knotting probability and the scaling behavior of self-avoiding polygons 
under a topological constraint
arXiv:1704.07510

\bibitem{PZC-16}
Plunkett Zirbel L and Chapman K 2016
Off-lattice random walks with excluded volume: a new method of generation,
proof of ergodicity and numerical results
{\em J. Phys.} A: {\em Math. Theor.} {\bf 49} 135203

\bibitem{VL-94}
Vega C and Lago S 1994
A fast algorithm to evaluate the shortest distance between rods
{\em Computers Chem.} {\bf 18} 55-59

\bibitem{RC-libro}
Rubinstein M and Colby R H C 2003
{\em Polymer Physics}
(Oxford: Oxford University Press)

\bibitem{ZSF-53}
Zimm B H, Stockmayer W H and Fixman M 1953
Excluded Volume in Polymer Chains
{\em J. Chem. Phys.} {\bf 21} 1716

\bibitem{Yamakawa-71}
Yamakawa H 1971 {\em Modern Theory of Polymer Solutions} \/
(New York: Harper--Row)

\bibitem{CMP-08}
Caracciolo S, Mognetti B M and Pelissetto A 2008
Two-parameter model predictions and $\theta$-point crossover for
linear-polymer solutions
{\em J. Chem. Phys.} {\bf 128} 065104

\bibitem{DPP-13-Thermal}
D'Adamo G, Pelissetto A and Pierleoni C 2013
Consistent coarse-graining strategy for polymer solutions 
in the thermal crossover from good to $\theta$ solvent
{\em J. Chem. Phys.} {\bf 139} 034901

\bibitem{MS-88}
Madras N and Sokal A D 1988 
The pivot algorithm: A highly efficient Monte Carlo method for the
self-avoiding walk
{\em J. Stat. Phys.} {\bf 50} 109

\bibitem{Kennedy-02}
Kennedy T 2002 
A faster implementation of the pivot algorithm for self-avoiding walks
{\em J. Stat. Phys.} {\bf 106} 407

\bibitem{Berry-66}
Berry G C 1966 
Thermodynamic and Conformational Properties of Polystyrene. I. Light-Scattering
Studies on Dilute Solutions of Linear Polystyrenes
{\em J. Chem. Phys.} {\bf 44} 4550

\bibitem{DPP-14-GFVT}
D'Adamo G, Pelissetto A and Pierleoni C 2014
Phase diagram of mixtures of colloids and polymers 
in the thermal crossover from good to $\theta$ solvent
{\em J. Chem. Phys.} {\bf 141} 024902

\bibitem{PH-05}
Pelissetto A and Hansen J-P 2005
Corrections to scaling and crossover from good- to $\theta$-solvent
regimes of interacting polymers
{\em J. Chem. Phys.} {\bf 122} 134904

\bibitem{KS-94}
Kr\"uger B and Sch\"afer L 1994 
Long polymer chains in good solvent: beyond the universal limit
{\em J. Phys. I (France)} {\bf 4} 757

\bibitem{Hall-80}
Hall C K 1980
Polymer scaling theories for general interaction potentials
{\em J. Chem. Phys.} {\bf 73} 1446

\bibitem{Isihara-50}
Isihara A 1950 
Determination of molecular shape by osmotic measurement
{\em J. Chem. Phys.} {\bf 18} 1446

\bibitem{IH-51}
Isihara A and Hayashida T 1951
Theory of high polymer solution. I. Second virial coefficient for rigid
ovaloids models
{\em J. Phys. Soc. Jpn.} {\bf 6}  40-45

\bibitem{Kihara-53}
Kihara T 1953 
Virial coefficients and models of molecules in gases
{\em Rev. Mod. Phys.} {\bf 25} 831

\bibitem{DJ-72}
Domb C and Joyce G S 1972
Cluster expansion for a polymer chain
{\em J. Phys.} C {\bf 5} 956-976

\bibitem{Louis}
Bolhuis P G, Louis A A and Hansen J P 2001 
Many-body interactions and correlations in coarse-grained descriptions of
polymer solutions
{\em Phys. Rev.} E {\bf 64} 021801

\bibitem{DPP-compressible}
D'Adamo G, Pelissetto A and Pierleoni C 2012
Polymers as compressible soft spheres
{\em J. Chem. Phys.} {\bf 136} 224905

\bibitem{Causo-02}
Causo M S 2002
Universal shape ratios for polymers grafted at a flat surface
{\em J. Chem. Phys.} {\bf 117} 6789

\bibitem{Sciutto-96}
Sciutto S J 1996 
The shape of self-avoiding walks
{\em J. Phys.} A: {\em Math. Gen.} {\bf 29} 5455

\bibitem{shape}
Zifferer G 1999
Monte Carlo simulation studies of the size and shape of linear and
star-branched polymers embedded in the tetrahedral lattice
{\em Macromol. Theory Simul.} {\bf 8} 433 \par
Zifferer G and Preusser W  2001
Monte Carlo simulation studies of the size and shape of ring polymers
{\em Macromol. Theory Simul.} {\bf 10} 397

\bibitem{SHG-16}
\v{S}krbi{\'c} T, Hoang T X and Giacometti A 2016
Effective stiffness and formation of secondary structures in a 
protein-like model
{\em J. Chem. Phys.} {\bf 145} 084904

\bibitem{BH-67}
Barker J A and Henderson D 1967
Perturbation theory and equation of state for
fluids: The square-well potential
{\em J. Chem. Phys.} {\bf 47} 2856 \par
Barker J A and Henderson D 1967
Perturbation theory and equation of state for fluids. II. A successful theory
of liquids
{\em J. Chem. Phys.} {\bf 47} 4714

\bibitem{JHGJM-12}
Jover J, Haslam A J, Galindo A, Jackson G and M\"uller E A 2012
Pseudo hard-sphere potential for use in continuous molecular-dynamics
simulation of spherical and chain molecules
{\em J. Chem. Phys.} {\bf 137} 144505

\bibitem{Pelissetto-08}
Pelissetto A 2008
Osmotic pressure and polymer size in semidilute polymer solutions
under good-solvent conditions
{\em J. Chem. Phys.} {\bf 129} 044901

\bibitem{star}
Likos C N, L\"{o}wen H, Watzlawek M, Abbas B, Jucknischke O,
Allgaier J and Richter D 1998
Star polymers viewed as ultrasoft colloidal particles
{\em Phys. Rev. Lett.} \textbf{80} 4450\par
Watzlawek M, Likos C N and L\"{o}wen H 1999
Phase diagram of star polymer solutions
{\em Phys. Rev. Lett.} \textbf{82} 5289\par
Laurati M, Stellbrink J, Lund R, Willner L, Richter D and
Zaccarelli E 2005
Starlike micelles with starlike interactions: A quantitative evaluation of
structure factors and phase diagram
{\em Phys. Rev. Lett.} {\bf 94} 195504\par
Menichetti R, Pelissetto A and Randisi F 2017
Thermodynamics of star polymer solutions:
A coarse-grained study
{\em J. Chem. Phys.} {\bf 146} 244908

\bibitem{BO-12}
Baiesi M and Orlandini E 2012
Universal properties of knotted polymer rings
{\em Phys. Rev.} E {\bf 86} 031805

\bibitem{Stigter-77}
Stigter D 1977
Interactions of highly charged colloidal cylinders with applications to
double-stranded DNA
{\em Biopolymers} {\bf 16} 1435

\bibitem{DC-09}
Dobrynin A V and Carrillo J-M Y 2009
Swelling of biological and semiflexible polyelectrolytes
{\em J. Phys. Condens. Matter} {\bf 21} 424112

\bibitem{BTSRDM-15}
Brunet A, Tardin C, Salom\'e L, Rousseau P, Destainville N
and Manghi M 2015
{\em Macromolecules} {\bf 48} 3641

\end{thebibliography}
\end{document}